\documentclass[prl,twocolumn,showpacs,preprintnumbers,amsmath,amssymb]{revtex4-1}

\usepackage{epsfig}% Include figure files
\usepackage{graphicx}% Include figure files
\usepackage{dcolumn}% Align table columns on decimal point
\usepackage{bm}% bold math

\newcommand{\EQ}{\begin{equation}}
\newcommand{\EN}{\end{equation}}
\newcommand{\ea}{\end{eqnarray}}
\newcommand{\ba}{\begin{eqnarray}}

\newcommand{\bear}{\begin{eqnarray}}
\newcommand{\ear}{\end{eqnarray}}

\begin{document}
\title{Diffusive charge transport in the gapped 1D Hubbard model at all finite temperatures}

%\date{}

\author{J. M. P. Carmelo}
\affiliation{Center of Physics of University of Minho and University of Porto, LaPMET, P-4169-007 Oporto, Portugal}
\affiliation{CeFEMA, Instituto Superior T\'ecnico, Universidade de Lisboa, LaPMET, Av. Rovisco Pais, P-1049-001 Lisboa, Portugal}
\author{P. D. Sacramento}
\affiliation{CeFEMA, Instituto Superior T\'ecnico, Universidade de Lisboa, LaPMET, Av. Rovisco Pais, P--1049-001 Lisboa, Portugal}

%\date{\today}

%%%%%%%%%%%%%%%%%%%%%%%%%%%%%%%%%%%%%%%%%%%%%%%%%%%%%%%%%%%%%%%%%%%%%%%%%%
%                              abstract                                  %
%%%%%%%%%%%%%%%%%%%%%%%%%%%%%%%%%%%%%%%%%%%%%%%%%%%%%%%%%%%%%%%%%%%%%%%%%%

\begin{abstract} 
Studies relying on hydrodynamic theory and Kardar-Parisi-Zhang (KPZ) scaling have found that 
in the one-dimensional Hubbard model spin and charge transport are for all temperatures $T>0$ 
anomalous superdiffusive at zero magnetic field, $h=0$, and zero 
chemical potential, $\mu =0$, respectively. However, this contradicts recent exact results 
that at very low temperature charge transport rather is normal diffusive. 
In this Letter we identify the mechanisms that control the different types of temperature dependence 
of the $h=0$ spin and $\mu =0$ charge transport and find that the latter is normal diffusive for {\it all}
finite temperatures $T>0$, in contrast to the hydrodynamic theory and KPZ scaling predictions.
\end{abstract}

\maketitle
%%%%%%%%%%%%%%%%%%%%%%%%%%%%%%%%%%%%%%%%%%%%%%%%%%%%%%%%%%%%%%%%%%%%%%%%%%
%                              body of paper                             
%%%%%%%%%%%%%%%%%%%%%%%%%%%%%%%%%%%%%%%%%%%%%%%%%%%%%%%%%%%%%%%%%%%%%%%%%%
%%%%%%%%%%%%%%%%%%%%%%%%%%%%%%%%%%%%%%%%%%%%%%%%%%%%%%%%%%%%%%%%%%%%%%%%%%

An interesting unsolved issue concerning the repulsive one-dimensional (1D) Hubbard model's \cite{Lieb_68,Takahashi_72,Martins_98}
type of charge transport at arbitrary finite temperature was brought about by recent results of Ref. \onlinecite{Carmelo_24}
that at zero chemical potential, $\mu =0$, it is normal diffusive for very low temperatures. 
Indeed, hydrodynamic theory and Kardar-Parisi-Zhang (KPZ) scaling studies had found that at zero chemical potential the
model's charge transport is for {\it all} temperatures $T>0$ anomalous 
superdiffusive \cite{Moca_23,Fava_20,Ilievski_18}.

For simplicity, we use $\alpha$-spin in $S_{\alpha}$ and $S_{\alpha}^z$ for charge/$\eta$-spin ($\alpha=\eta$) and spin ($\alpha=s$) and
denote $\mu$ and $2\mu_B h$ by $\mu_{\alpha}$ for $\alpha=\eta$ and $\alpha=s$, respectively. Here $\mu_B$ is 
the Bohr magneton and $h$ the magnetic field. 

The goal of this Letter is to clarify the unsolved issue concerning the type of charge transport of the 1D Hubbard model at 
the $h = \mu =0$ point for finite temperatures $T>0$. We find 
that charge transport is normal diffusive for {\it all} finite temperatures $T>0$, in contrast to the
predictions of hydrodynamic theory and KPZ scaling \cite{Moca_23,Fava_20,Ilievski_18}.
Our studies rely on the $\tau$- and $\alpha$-spin 
representation of the 1D Hubbard model in its full Hilbert space recently introduced in Ref. \onlinecite{Carmelo_25}.

Compounds for which transport experiments are suggested to search for the normal charge diffusive transport
at finite temperatures found for the 1D Hubbard model at zero chemical potential include all quasi-1D Mott-Hubbard insulators
 \cite{Du_24,Ono_04}. If quasi-1D systems such as nanoribbons \cite{Popple_23} could effectively mimic the 1D Hubbard model at zero chemical potential
through careful tuning of interactions and dimensionality, they could be used to experimentally identify that model's
finite-temperature normal diffusive transport. The systems degrade or lose coherence at high temperatures due to thermal 
vibrations or substrate effects, so that for actual compounds and systems our results are valid at temperature scales at which they  
effectively mimic the 1D Hubbard model.

Finite-temperature spin and/or charge transport in integrable correlated models
is actually an intensely debated fundamental quantum problem. The recent use of hydrodynamic theory, 
KPZ scaling, and other methods to study it is indeed receiving considerable attention, giving rise to 
a large number of publications 
\cite{Carmelo_24,Moca_23,Fava_20,Ilievski_18,Carmelo_25,Ljubotina_17,Medenjak_17,Nardis_18,Gopalakrishnan_19,Ljubotina_19A,Ljubotina_19,Ye_20,Nardis_20,Nardis_21,Ilievski_21,Krajnik_22,Ye_22,Nardis_23A}.

The 1D Hubbard model describes correlations of $N = N_{\uparrow} + N_{\downarrow}$ electrons in a lattice of length $L$ with $N_a$ sites
and densities $m_{\eta z} = (N_a-N)/L$ and $m_{s z} = (N_{\uparrow} - N_{\downarrow})/L$.
We consider the thermodynamic limit, its Hamiltonian under periodic boundary conditions reading at $\mu_{\alpha} =0$
\cite{Lieb_68,Takahashi_72,Martins_98,Carmelo_17A},
\begin{eqnarray}
\hat{H} & = & -t\sum_{\sigma, j}\left[c_{j,\sigma}^{\dag}\,c_{j+1,\sigma} + 
{\rm h.c.}\right] + U\sum_{j}\hat{\rho}_{j,\uparrow}\hat{\rho}_{j,\downarrow} \, .
\label{H}
\end{eqnarray}
Here $c_{j,\sigma}^{\dag}$ creates one electron of spin projection $\sigma$ at site $j$,
$\hat{\rho}_{j,\sigma}= (\hat{n}_{j,\sigma}-1/2)$, and $\hat{n}_{j,\sigma}=c_{j,\sigma}^{\dag}\,c_{j,\sigma}$.
We use natural units in which the Planck constant, electronic charge, and lattice spacing
are equal to 1, so that $N_a=L$. Our results refer to $u=U/4t >0$.

Within the {\it $\tau$- and physical $\alpha$-spin representation} \cite{Carmelo_25,SM} the $4^L$ 
energy eigenstates that span the whole Hilbert space refer to $4^L$ irreducible representations
of the model $[SU (2)\times SU(2)\times U(1)]/Z_2^2$ global symmetry \cite{Carmelo_25,Carmelo_10}.
It contains the $\tau$-translational $U(1)$ symmetry the generator of which has eigenvalue 
$S_{\tau} = {1\over 2}L_{\eta} = {1\over 2}(L-L_{s})$. Here $L_{\eta}$ and $L_{s}$ are the
numbers of physical $\eta$-spins and physical spins, respectively, such that $L_{\eta} + L_{s} = L$ \cite{Carmelo_25,SM}.

For the numbers $N_{\alpha}$ of {\it unpaired physical $\alpha$-spins} and ${\cal{N}}_{\alpha}$
of {\it paired physical $\alpha$-spins} such that $L_{\alpha} = N_{\alpha} + {\cal{N}}_{\alpha}$ for $\alpha = s,\eta$,
%CORR the numbers $M_{\alpha n}$ of $\alpha$-spin configurations called {\it $\alpha n$-pairs} where $n=1,...,\infty$ is their number of 
%physical $\alpha$-spins pairs ADDED
the numbers $M_{\alpha n}$ of $\alpha$-spin configurations called {\it $\alpha n$-pairs} where $n=1,...,\infty$ is their number of 
physical $\alpha$-spins pairs, and the discrete $\tau$-band and $\alpha n$-band momentum values
$q_j = {2\pi\over L} I_j^{\tau}$ and $q_j = {2\pi\over L} I_j^{\alpha n}$
%CORR of each $n=1,...,\infty$ $\alpha n$-pair branch ADDED
of each $n=1,...,\infty$ $\alpha n$-pair branch, respectively,
see Refs. \onlinecite{Carmelo_25,SM}. The latter have Pauli-like occupancies: The $\tau$-band and $\alpha n$-band momentum distributions read 
(i) $N_{\tau} (q_j) = 1$ and $M_{\alpha n} (q_j) = 1$ and (ii) $N_{\tau} (q_j) = 0$ and $M_{\alpha n} (q_j) = 0$
for (i) occupied and (ii) unoccupied $q_j$'s, respectively. We also define corresponding $\tau$-hole
and $\alpha n$-hole distributions $N_{\tau}^h (q_j) = 1 - N_{\tau} (q_j)$ and
$M_{\alpha n}^h (q_j) = 1 - M_{\alpha n} (q_j)$, respectively. As reported in Refs. \onlinecite{Carmelo_25,SM}, one
can use a $\tau$-band momentum-rapidity-variable distribution ${\bar{N}}_{\tau} (k)$ where $k\in [-\pi,\pi]$
and $\alpha n$-bands rapidity-variable distribution ${\bar{M}}_{\alpha n} (\Lambda)$ where $\Lambda\in [-\infty,\infty]$ 
that store exactly the same information as those in terms of the $q_j$'s. The relation between the two
notations involves the  $\tau$-band momentum-rapidity
function $k = k (q_j)$ and  $\alpha n$-bands rapidity functions $\Lambda_{\alpha n} (q_j)$
where $\alpha = s,\eta$ and $n=1,...,\infty$ \cite{Carmelo_25,SM}. 

The {\it $\alpha$-spin carriers} of {\it all} $S_{\alpha}>0$ energy eigenstates are the $N_{\alpha} = 2S_{\alpha}$
unpaired physical $\alpha$-spins \cite{Carmelo_25,SM}.
The $N_{\alpha,\pm 1/2}= S_{\alpha} \pm S_{\alpha}^z$ $\alpha$-spin carriers of projection $+1/2$ and $-1/2$
have couplings of opposite sign to a $\alpha$-spin vector potential \cite{Shastry_90,Carmelo_18}. 
They carry $\alpha$-spin elementary currents $j_{\alpha,\pm 1/2}$ 
such that $j_{\alpha,+ 1/2} = - j_{\alpha,-1/2}$ that play a key role in our studies 
and the expressions of which are known for all $S_{\alpha}>0$ energy eigenstates \cite{Carmelo_25,SM}.

The real part of the model's $\alpha$-spin conductivity reads,
\begin{equation}
\sigma_{\alpha} (\omega,T) = 2\pi D^z_{\alpha} (T)\delta (\omega) +\sigma_{\alpha, {\rm reg}} (\omega,T) 
\hspace{0.20cm}{\rm where}\hspace{0.20cm}\alpha = \eta,s \, .
\label{sigma}
\end{equation}
For some of its limiting behaviors for charge \cite{Carmelo_04}, see Ref. \onlinecite{SM}.
It is well established that at $\mu_{\alpha}=0$ $\alpha$-spin transport is not ballistic \cite{Ilievski_17}.

At the $h = \mu = 0$ point the $\alpha$-spin-diffusion constant can be expressed as \cite{Carmelo_25},
\begin{eqnarray}
D_{\alpha} (T) & = & C_{\alpha} (T)\,\Pi_{\alpha} (T) = C_{\alpha} (T)\,L\,\langle\vert j_{\alpha,\pm 1/2}\vert^2\rangle_{m_{{\bar{\alpha}} z},T} 
\nonumber \\
&& {\rm where}\hspace{0.20cm}\bar{\eta} = s \hspace{0.20cm}{\rm and}\hspace{0.20cm}\bar{s} = \eta \, ,
\label{DproptoOmega}
\end{eqnarray}
and the thermal expectation value reads,
\begin{eqnarray}
&& \langle\vert j_{\alpha,\pm 1/2}\vert^2\rangle_{m_{{\bar{\alpha}} z},T} = {\Pi_{\alpha} (T)\over L} 
\nonumber \\
&& = \sum_{z= -1}^{1}
\sum_{2S_{\alpha}=1}^{L_{\alpha}}
\sum_{S_{\alpha}^z=-S_{\alpha}}^{S_{\alpha}}
\sum_{2S_{\bar{\alpha}}=\vert 2S_{\bar{\alpha}}^z\vert}^{L_{\bar{\alpha}}} \sum_{l_{\rm r}^{u}} 
p_{l_{\rm r}^{u},S_{\tau},S_{\alpha},S_{\alpha}^z,S_{\bar{\alpha}},S_{\bar{\alpha}}^z}
\nonumber \\
&& \hspace{1.5cm}\times \vert j_{\alpha,\pm 1/2}\vert^2 \, .
\label{jz2TD}
\end{eqnarray}

Here the sum $\sum_{z= (L_{\eta}-L_s)/L = -1}^{1}$ is equivalent to $\sum_{2S_{\tau}=0}^{L}$,
$S_{\bar{\alpha}}^z = 0$ and $m_{{\bar{\alpha}} z} =0$ in our $\mu = h =0$ case, and
the sum $\sum_{S_{\alpha}^z=-S_{\alpha}}^{S_{\alpha}}$ runs over states outside the Bethe-ansatz subspace \cite{SM}. 
The quantities $p_{l_{\rm r}^{u},S_{\tau},S_{\alpha},S_{\alpha}^z,S_{\bar{\alpha}},S_{\bar{\alpha}}^z}$ are 
the Boltzmann weights where $l_{\rm r}^{u}$ denotes all quantum numbers other than 
$S_{\tau}$, $S_{\alpha}$, $S_{\alpha}^z$, $S_{\bar{\alpha}}$, and $S_{\bar{\alpha}}^z$
needed to specify an energy eigenstate and the coefficient $C_{\alpha} (T)$ is finite for $T>0$ \cite{Carmelo_25,SM}. 
Such a finiteness implies that the thermal expectation value $\langle\vert j_{\alpha,\pm 1/2}\vert^2\rangle_{m_{{\bar{\alpha}} z},T} = 
\Pi_{\alpha} (T)/L$ determines for $T>0$ whether $\alpha$-spin transport is normal diffusive or anomalous superdiffusive, respectively.
\begin{figure*}
\includegraphics[width=0.495\textwidth]{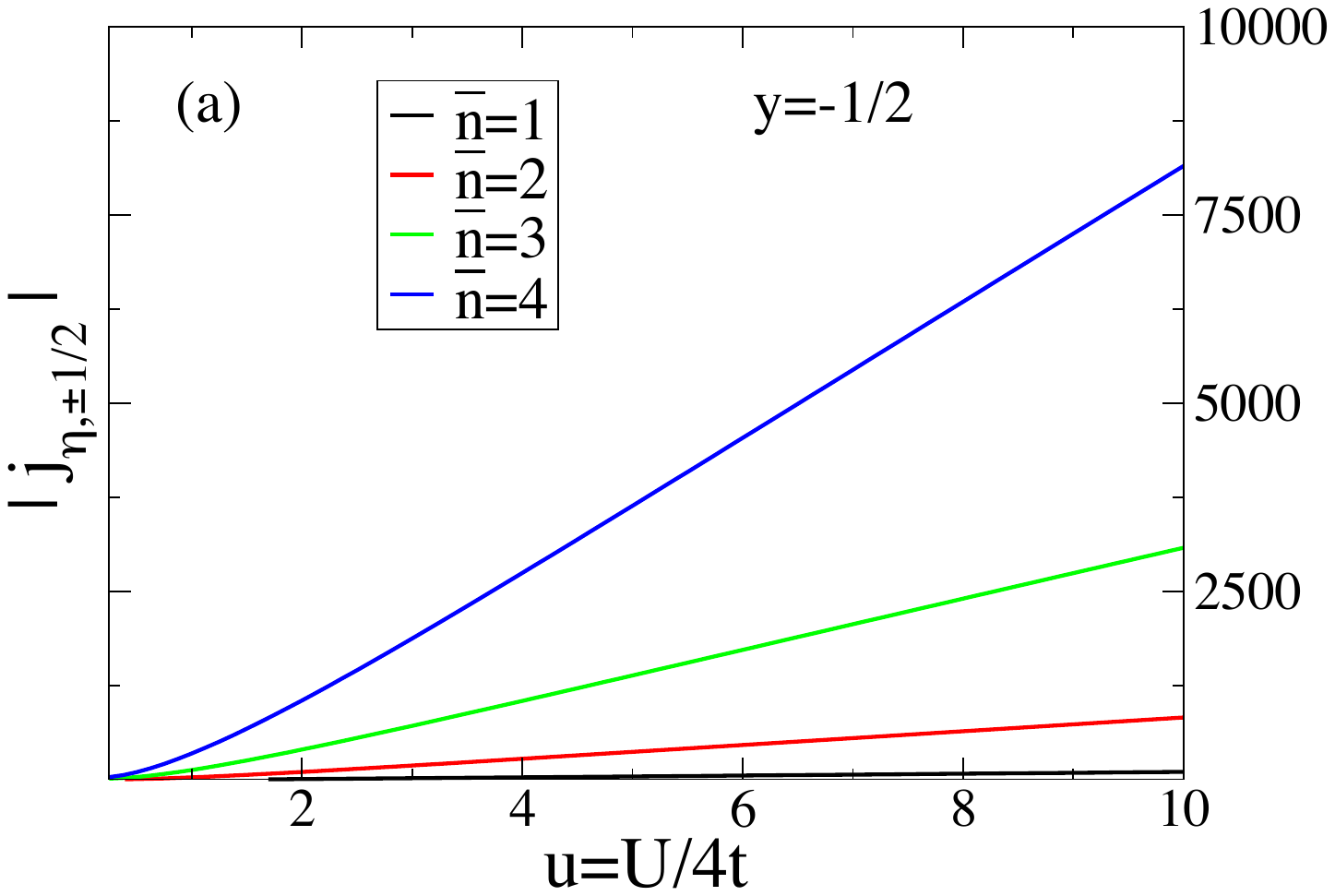}
\includegraphics[width=0.495\textwidth]{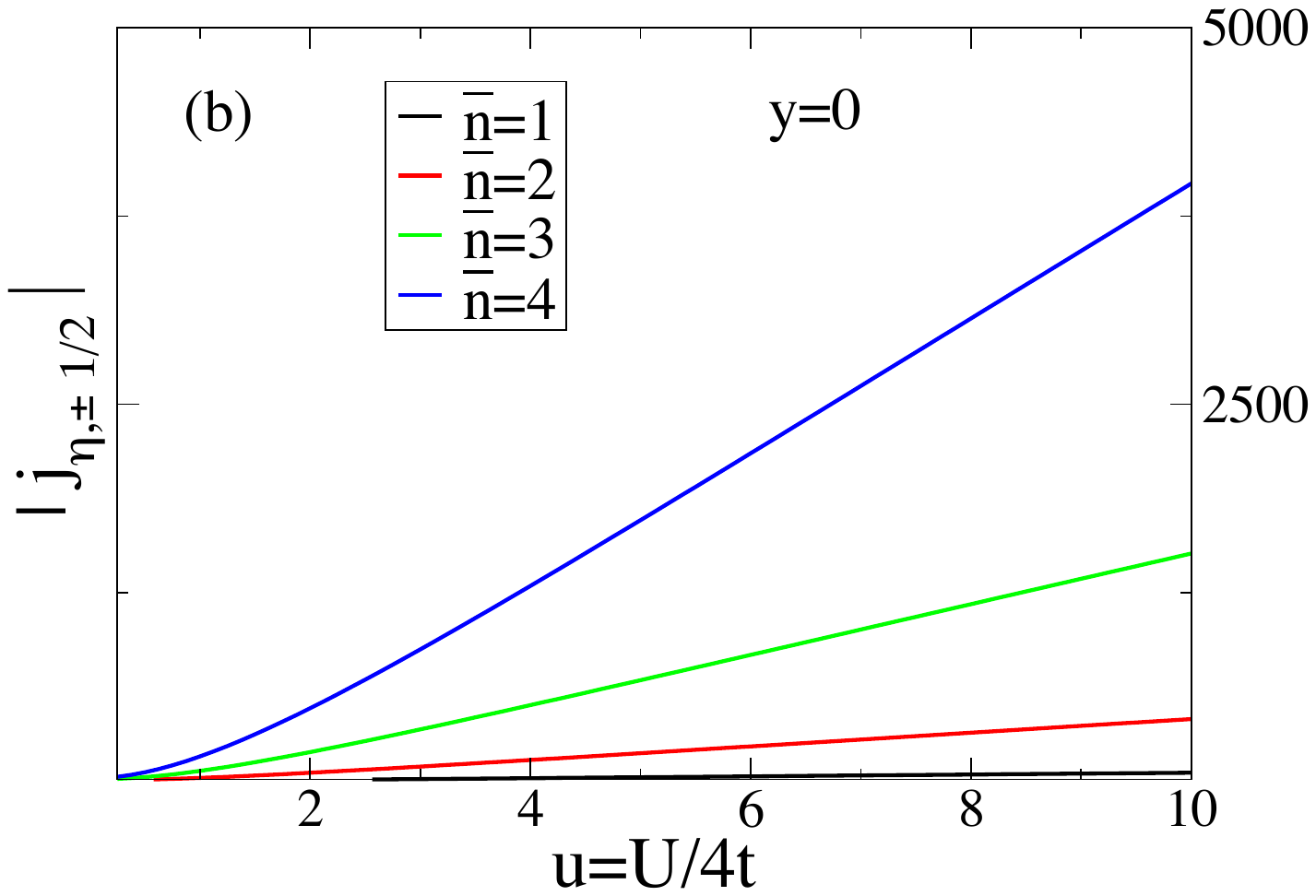}
\includegraphics[width=0.495\textwidth]{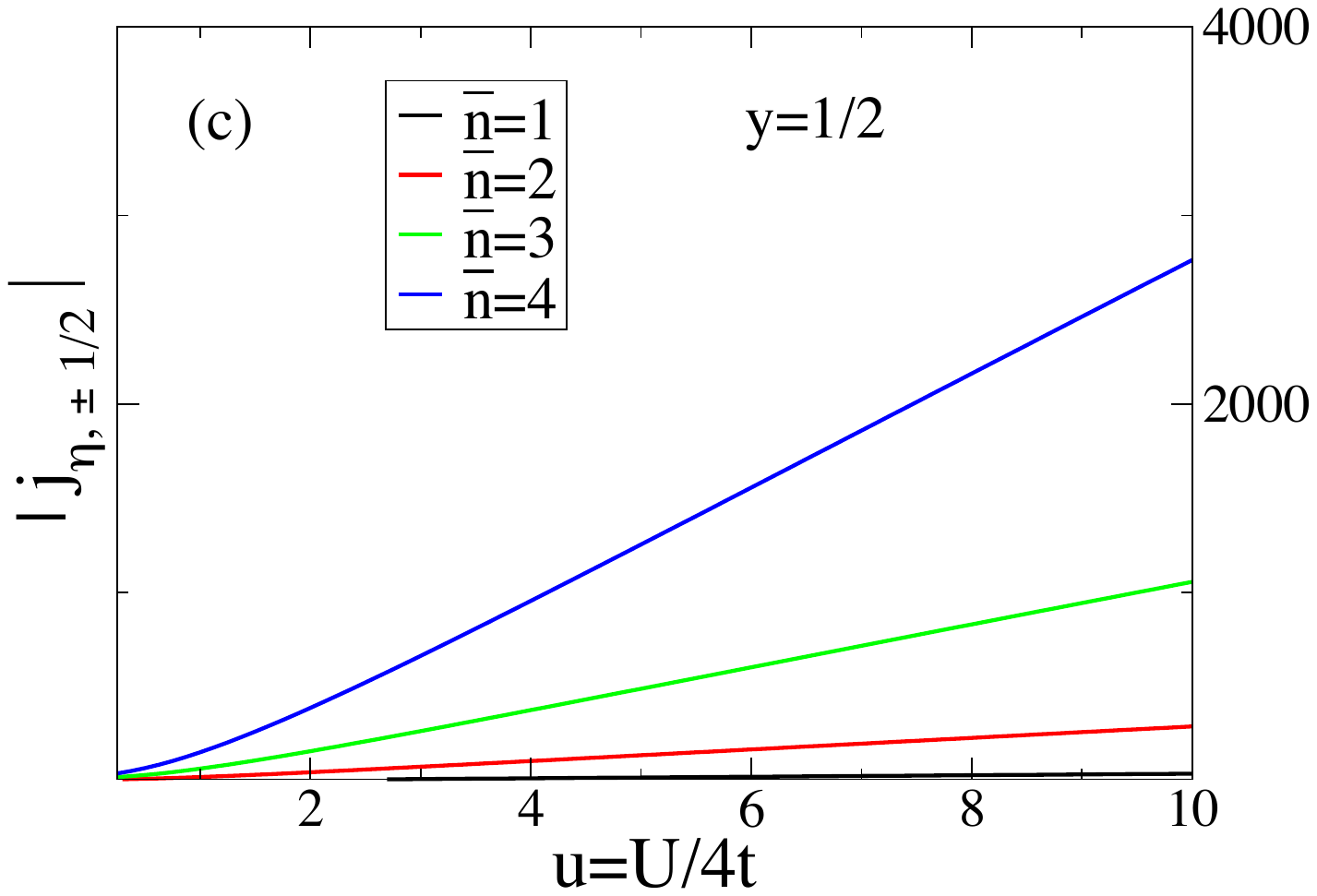}
\includegraphics[width=0.495\textwidth]{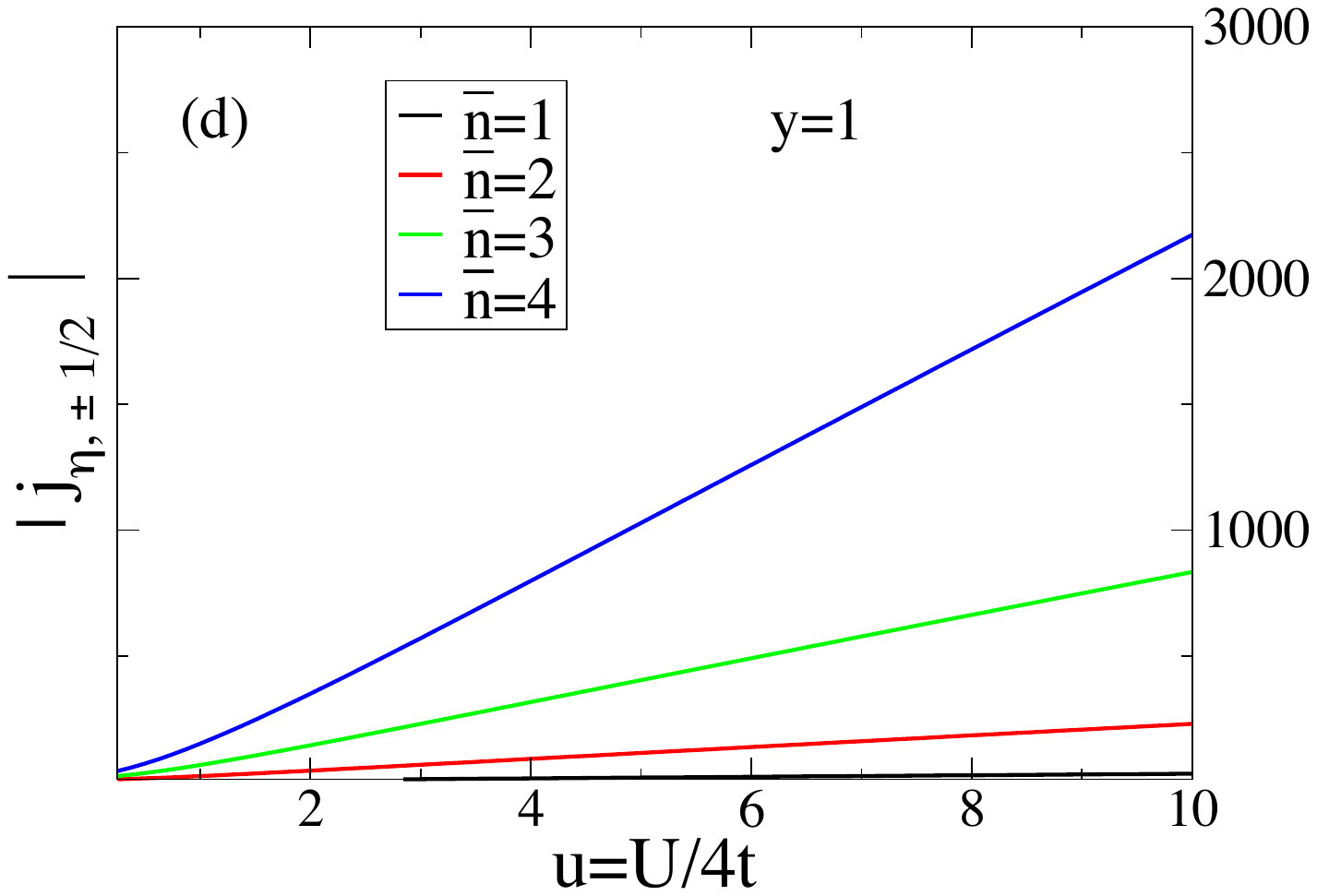}
\caption{The absolute values $\vert j_{\eta, \pm 1/2}(\bar{n})\vert$ of the charge elementary current carried by one 
charge carrier for $\bar{n}$-states, Eq. (\ref{setancurrentsG}) for $\alpha = \eta$, with $\bar{n}=1,2,3,4$, as a function of $u=U/4t$ for (a) $y = -1/2$,
(b) $y = 0$, (c) $y = 1/2$, and (d) $y = 1$. For $\bar{n}\rightarrow\infty$, $\vert j_{\eta, \pm 1/2}(\bar{n})\vert = \infty$ for $u>0$.}
\label{figure1PRL}
\end{figure*}

We use the notations $\alpha = s,\eta$ and $\bar{\alpha} = \eta,s$, Eq. (\ref{DproptoOmega}).
Most energy eigenstates at the $h = \mu = 0$ point
are of class (A) for which $\vert j_{\alpha,\pm 1/2}\vert^2$ is of the order of $1/L$ and vanishes in the thermodynamic limit. 
In addition, there is a finite density of important energy eigenstates of class (B) for which $\vert j_{\alpha,\pm 1/2}\vert^2$ is finite
or even infinite. As justified below, the type of 
excitation energy density spectrum of such states of class (B) determines whether $T>0$ $\alpha$-spin transport is  normal diffusive or 
anomalous superdiffusive. 

The use of the general expression of the $\alpha$-spin elementary currents $j_{\alpha,\pm 1/2}$ \cite{SM} justifies why
the square of their absolute values are for the states of class (A) of the order of $1/L$: Most $\alpha n$-bands occupancy
configurations that generate such energy eigenstates have both occupied and unoccupied discrete momentum
values for $q_j <0$ and $q_j >0$, respectively. This implies that many of the contributions to the $\alpha$-spin elementary currents
from $q_j <0$ and $q_j >0$ occupancies cancel each other. 

In contrast, the states of class (B) have compact occupancies 
of $\alpha n$-holes and $\alpha n$-pairs with momentum values $q_j$ of opposite
sign, respectively, that provide finite contributions to the $\alpha$-spin elementary currents. 
Those with largest finite $\alpha$-spin elementary currents have in rapidity space 
$\alpha n$-bands the $\alpha n$-holes of which have compact occupancy for $\Lambda <0$ and the $\alpha n$-pairs
of which have compact occupancy for $\Lambda >0$, respectively, or vice versa. 
In addition to other states of class (B) \cite{SM}, here we consider those that have the largest absolute values 
$\vert j_{\alpha,\pm 1/2}\vert$. Their rapidity-variable distributions are given by,
\begin{eqnarray}
{\bar{N}}_{\tau} (k) & = & 1 \hspace{0.20cm}{\rm for}\hspace{0.20cm}k\in [\pi y,\pi]
\nonumber \\
& = & 0\hspace{0.20cm}{\rm for}\hspace{0.20cm}k\in [-\pi,\pi y]
\nonumber \\
{\bar{M}}_{\alpha n} (\Lambda) & = & 1 \hspace{0.20cm}{\rm for}\hspace{0.20cm}\Lambda\in [0,\infty]
\hspace{0.20cm}{\rm and}\hspace{0.20cm}n = 1,...,\bar{n}_{\alpha}
\nonumber \\
& = & 0\hspace{0.20cm}{\rm for}\hspace{0.20cm}
\Lambda\in [-\infty,0]\hspace{0.20cm}{\rm and}\hspace{0.20cm}n = 1,...,\bar{n}_{\alpha}
\nonumber \\
{\bar{M}}_{\alpha n} (\Lambda) & = & 0\hspace{0.20cm}{\rm for}\hspace{0.20cm}n > \bar{n}_{\alpha} \, ,
\label{NNstates1}
\end{eqnarray}
both for $\alpha =s$ and $\alpha =\eta$. Here $y$ is defined in terms of the $\tau$-band 
momentum-rapidity function $k (q)$ as $y = y (z) = k (\pi z + q^{\Delta})/\pi\in [-1,1]$. It corresponds to a shifted $\tau$-band momentum 
$q - q^{\Delta} = \pi z = \pi(L_{\eta} - L_s)/L \in [-\pi,\pi]$
%CORR \cite{Carmelo_25,SM} ADDED
\cite{Carmelo_25,SM}. We call $z (y)$ the inverse function of $y (z)$. It
obeys the relations $z (\pm 1) = \pm 1$ and $z (0) = 0$. Except for $u\rightarrow\infty$
when $z (y) = y$ for $y \in [-1,1]$, $z (y) \neq y$ for $y \neq 0,\pm 1$.
\begin{figure*}
\includegraphics[width=0.495\textwidth]{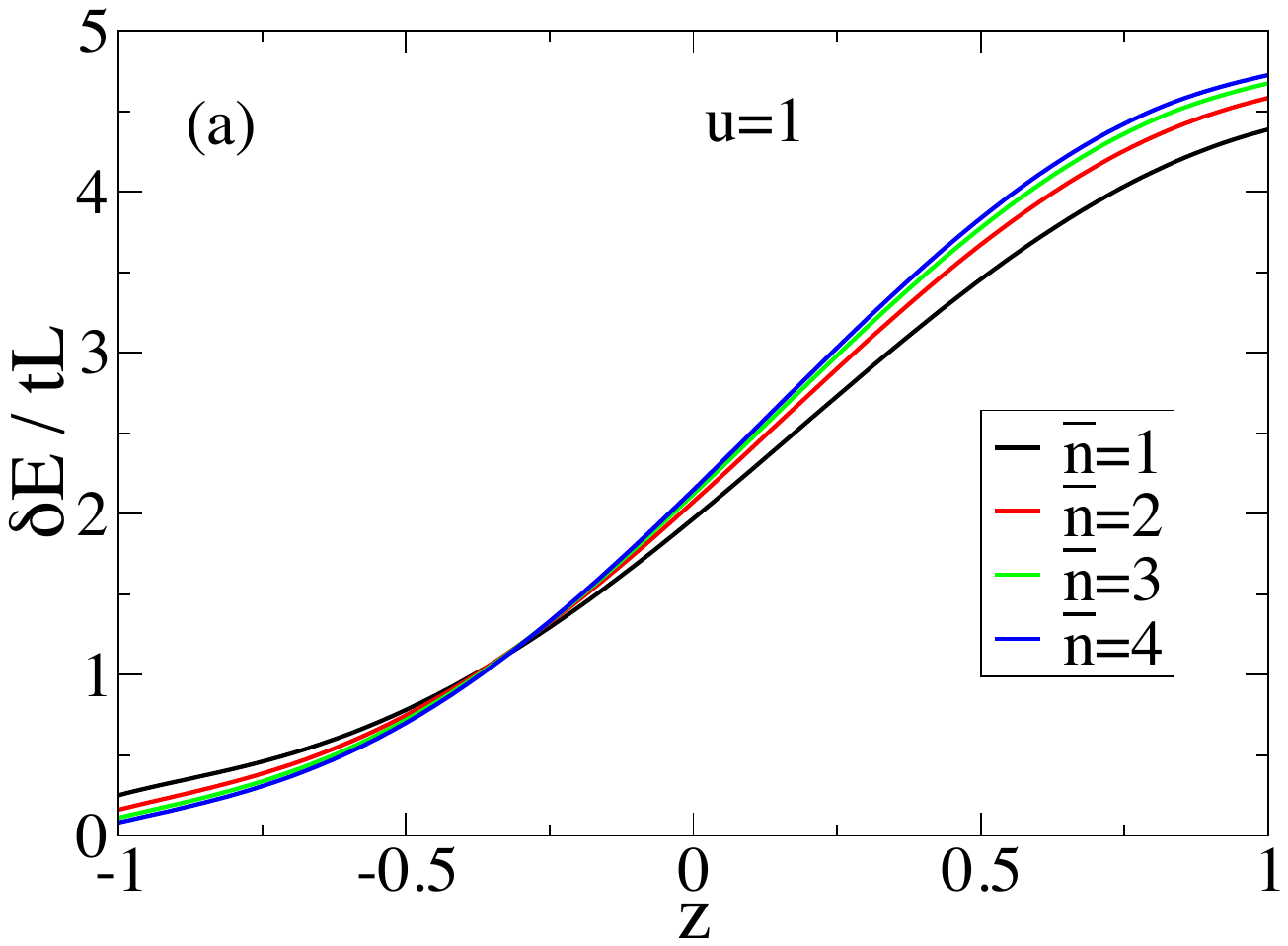}
\includegraphics[width=0.495\textwidth]{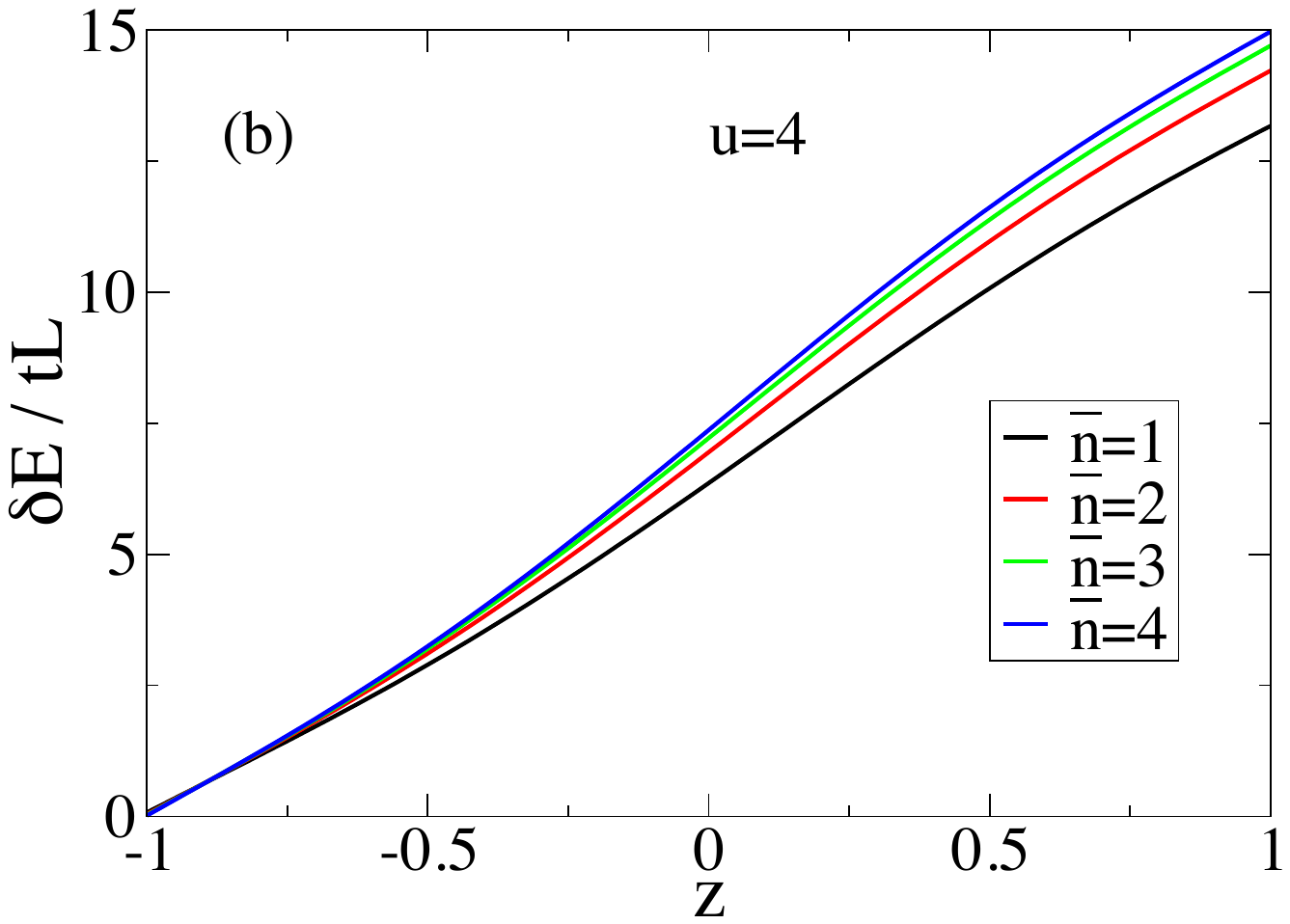}
\includegraphics[width=0.495\textwidth]{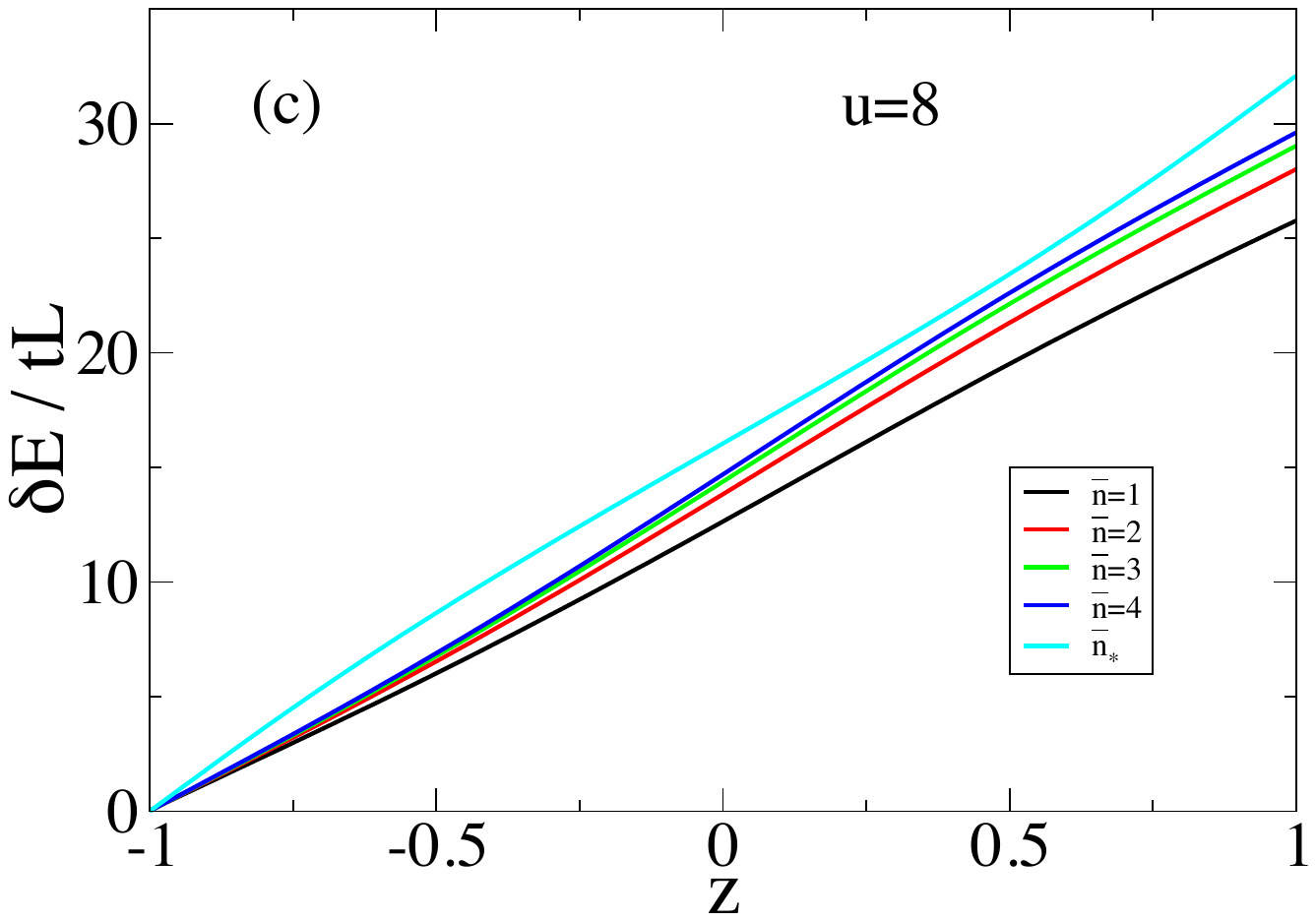}
\includegraphics[width=0.495\textwidth]{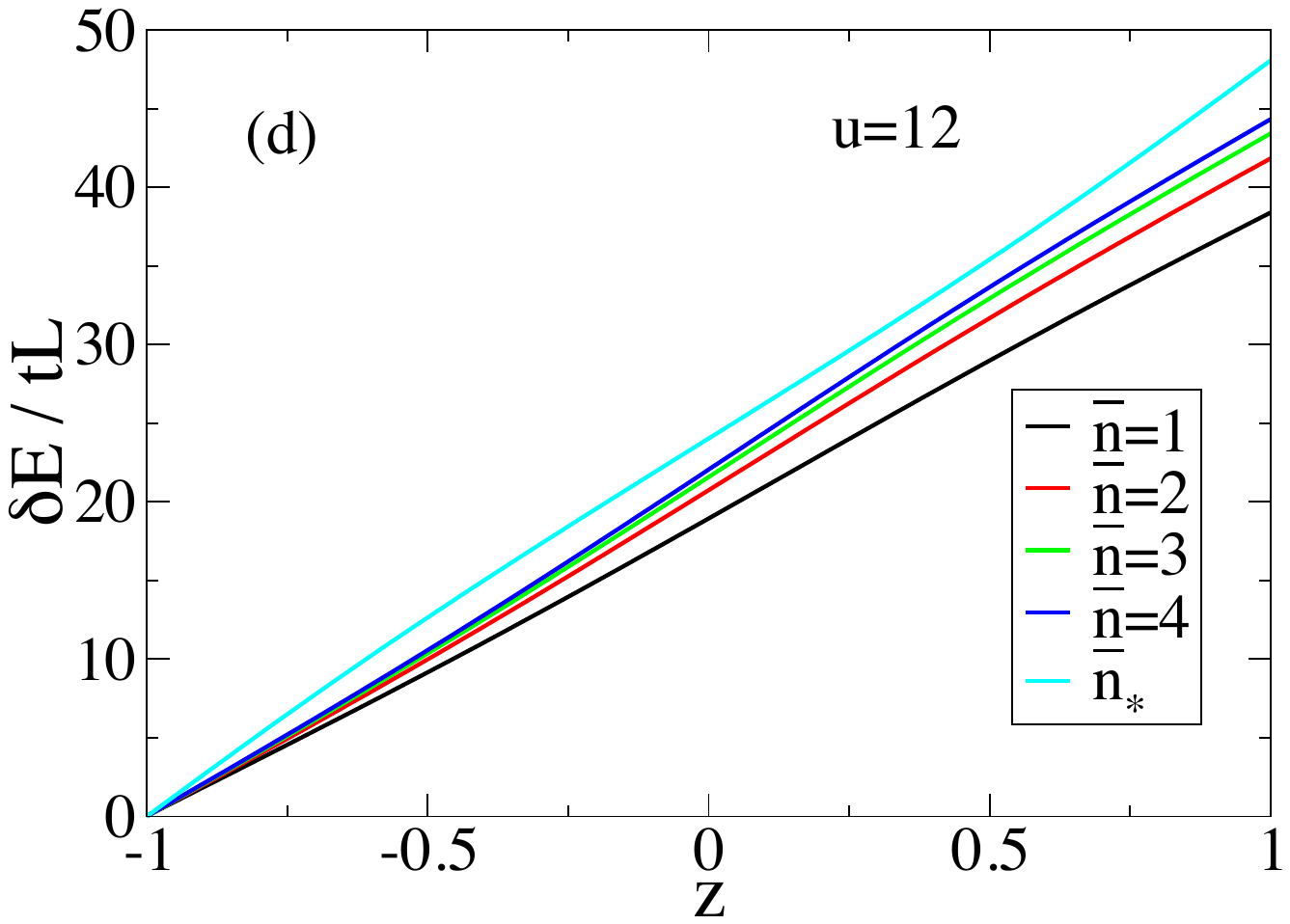}
\caption{The excitation energy density in units of transfer integral $t$ as a function of
$z = (L_{\eta} - L_z)/L \in [-1,1]$, Eq. (\ref{Enstates}), for (a) $u=1$ and (b) $u=4$ and $\bar{n}=1,2,3,4$
and for (c) $u=8$ and (d) $u=12$ and $\bar{n}=1,2,3,4,\bar{n}_*$ where $\bar{n}_*$
denotes $\bar{n}\rightarrow\infty$.}
\label{figure2PRL}
\end{figure*}

For simplicity, we consider $\bar{n}$ states for which $\bar{n} = \bar{n}_{s} = \bar{n}_{\eta}$, yet similar results are found for
states for which $\bar{n}_{s} \neq \bar{n}_{\eta}$. The $q_j$'s compact occupancies corresponding to those given in 
Eq. (\ref{NNstates1}) are provided in Ref. \onlinecite{SM}.  The use of the distributions of Eq. (\ref{NNstates1}) in the general expressions 
of the elementary currents $j_{\alpha,\pm 1/2}$ provided in that reference, gives for $\bar{n}$ states,
\begin{equation}
j_{\alpha,\pm 1/2}  = \pm {t\over\pi}\,(1 + \cos \pi y)\,{L\over L_{\alpha}}
\pm \iota_{\alpha}\sum_{n=1}^{\bar{n}}n{L\over M_{\alpha n}^h} j_{\alpha n} \, ,
\label{setancurrentsG}
\end{equation}
where $\alpha = s,\eta$, $\iota_s = 1$, $\iota_{\eta} = - 1$, and the expressions of $j_{\alpha n}$ and 
$L/M_{\alpha n}^h >1$ associated with $M_{\alpha n}^h/L <1$ are given in Ref. \onlinecite{SM}. 

The expression for $\bar{n}$ states of the concentration of $\alpha$-spin carriers, $2S_{\alpha}/L$, is also provided in that reference,
where it is plotted as a function of $u$ for $\alpha = \eta$ and $\bar{n} =1,2,3,4$.
It is found that $\bar{n}$ is finite for $u>0$ when the concentration of $\alpha$-spin carriers 
is finite in the $u\rightarrow\infty$ limit. On the other hand, we find that $\bar{n}\rightarrow\infty$ 
for $u>0$ when the number of $\alpha$-spin carriers $2S_{\alpha} = 1,2,3,...$
is finite and thus $2S_{\alpha}/L$ vanishes for $u\rightarrow\infty$ \cite{SM}. The absolute values
of the elementary currents $j_{\alpha,\pm 1/2}$ are for $u>0$
finite and diverge, respectively, in these two cases \cite{SM}. 

Since there is a finite density of states of class (B), $\Pi_{\alpha} (T)$ in Eq. (\ref{jz2TD})
and thus the $\alpha$-spin diffusion constant, Eq. (\ref{DproptoOmega}), diverge in the
thermodynamic limit, $L\rightarrow\infty$, for values of $k_B T$ corresponding to the
excitation energies associated with their energy eigenvalues. 

Finite absolute values of $\eta$-spin elementary currents, Eq. (\ref{setancurrentsG}) for $\alpha = \eta$, 
which diverge for $u>0$ when $\bar{n}\rightarrow\infty$, are plotted 
in Fig. \ref{figure1PRL} as a function of $u$ for $\bar{n}$ states with $\bar{n} = 1,2,3,4$ for 
(a) $y=-1/2$, (b) $y = z = 0$, (c) $y= + 1/2$, and (d) $y = z = +1$. 
Also the spin elementary currents of $\bar{n}$ states not plotted in the figure are finite for $\bar{n} = 1,2,3,4$ and $u>0$. 
While $z (0) = 0$ and $z (1) = 1$, $z  (y) = (L_{\eta} - L_s)/L$ becomes at $y=\pm 1/2$ closer to $\pm 1/2$ as $u$ increases. 
For instance, at $u=U/4t=1$ we have that $z (-1/2) \approx - 0.57$ for $\bar{n} = 1$, $z (-1/2) \approx - 0.58$ 
for $\bar{n} = 2,3,4$, and $z (1/2) \approx 0.63$ for $\bar{n} = 1,2,3,4$. 

The excitation energy density of $\bar{n}$ states is in units of the transfer integral $t$ given by,
\begin{eqnarray}
&& {\delta E\over L t} = {\sin \pi y\over \pi} + 4u\,{S_{\eta}\over L}
+ 4\int_0^{\infty}d\omega {J_0 (\omega)J_1 (\omega)\over\omega (1 + e^{2\omega u})}
\nonumber \\
&& - {1\over u}\sum_{\alpha = s,\eta}\sum_{n=1}^{\bar{n}}{n\over\pi^2}
\int_0^{\infty}d\Lambda\,2\pi\sigma_{\alpha n} (\Lambda)
\int_{\pi y}^{\pi}dk\,{\cos^2 k\over 1 + \Bigl({\sin k -  \Lambda\over n u}\Bigr)^2} 
\nonumber \\
&& + {4\over\pi}\sum_{n=1}^{\bar{n}}\int_0^{\infty}d\Lambda\,
2\pi\sigma_{\eta n} (\Lambda)\,{\rm Re}\left\{\sqrt{1 - (\Lambda - i n u)^2}\right\} \, .
\label{Enstates}
\end{eqnarray}
Here $J_0 (\omega)$ and $J_1 (\omega)$ are Bessel functions and the functions
$2\pi\sigma_{\alpha n} (\Lambda)$ are defined by integral equations \cite{SM}. The 
energy density $\delta E/L t$, Eq. (\ref{Enstates}), is plotted 
in Fig. \ref{figure2PRL} as a function of
$z = (L_{\eta} - L_s)/L \in [-1,1]$ for (a) $u=1$, (b) $u=4$, (c) $u=8$, and (d) $u=12$.
For $u=1$ and $u=4$ it is plotted for $\bar{n}$ states with $\bar{n} = 1,2,3,4$.
For $u=8$ and $u=12$ it is plotted for $\bar{n}$ states with $\bar{n} = 1,2,3,4,\infty$.

In the case of finite occupancy in an infinite number $\bar{n}\rightarrow\infty$ of $\alpha n$-bands, the evaluation of the 
energy density, Eq. (\ref{Enstates}), is for $u>0$ a technically complex problem. However, from manipulations
of Eq. (\ref{Enstates}) and the Bethe-ansatz equations, we find that for 
$\bar{n}\rightarrow\infty$ and the whole $u>0$ range the 
energy density $\delta E/L t$ vanishes at $z=-1$, being finite for $z > -1$. We could calculate $\delta E/L t$ for $\bar{n}\rightarrow\infty$
and large $u$ up to the order of $4t^2/U = t/u$, being plotted in Fig. \ref{figure2PRL} as a function of $z\in [-1,1]$ 
for (c) $u=8$ and (d) $u=12$.

Importantly, in the thermodynamic limit a finite 
energy density $\delta E/L t$ implies that the excitation energy $\delta E$ is
infinite. As shown in Fig. \ref{figure2PRL}, for $\bar{n}$ states with $\bar{n} = 1,2,3,4$ the energy
density is finite for $z\in [-1,1]$ at $u=1$ whereas for $u=4,8,12$ it vanishes only at $z=-1$ when $L_s = L$ and $L_{\eta} =0$
and there are no $\eta$-spins. On the other hand, for $\bar{n}\rightarrow\infty$ states the 
energy density vanishes for all $u>0$ values at $z=-1$ and is finite for all remaining $z\in ]-1,1]$ values. 

As shown in Eq. (\ref{jz2TD}), the $\alpha$-spin diffusion constant, Eq. (\ref{DproptoOmega}), has contributions
from sets of $S_{\alpha}>0$ energy eigenstates with different $z\in [-1,1]$ values. As discussed above, it has contributions 
from a finite density of states of class (B) with vanishing 
energy densities and thus finite energies only at $z = -1$, when the quantum problem is a spin-only 
$L_s = L$ system. Such energy densities vanish at $z = -1$ for the whole $u>0$ range in the case of $\bar{n}$ states with large 
$\bar{n}$ values.

For large $u$ up to the order of $4t^2/U = t/U$, $z=-1$ actually
refers to the spin-$1/2$ $XXX$ chain. The vanishing 
energy density $\delta E/L t=0$ at the spin-only point, $z=-1$, refers for $u>0$ to all finite values 
of the excitation energy $\delta E\in ]0,\infty]$ of large-$\bar{n}$ states. As shown in Fig. (\ref{figure2PRL}), for $u\geq 4$ this also applies to $\bar{n}$ states
with $\bar{n}=1,2,3,4$. Since $\Pi_{s} (T)$ in Eq. (\ref{DproptoOmega}) for $\alpha = s$ is infinite for such states
for which $z=-1$ and $\delta E \in ]0,\infty]$, this is why the spin diffusion constant of the 1D Hubbard model and 
spin-$1/2$ $XXX$ chain diverges for $k_B T \in ]0,\infty]$, as also found by hydrodynamic theory and KPZ scaling
\cite{Moca_23,Fava_20,Ilievski_18,Ljubotina_17,Medenjak_17,Nardis_18,Gopalakrishnan_19,Ljubotina_19A,Ljubotina_19,Ye_20,Nardis_20,Nardis_21,Ilievski_21,Krajnik_22,Ye_22,Nardis_23A}.

On the other hand, for $z>-1$ when both $L_s$ and $L_{\eta}$ are finite and particularly at $z =1$ when 
$L_s = 0$ and $L_{\eta} = L$, the energy density $\delta E/L t$ is {\it always} finite for $u>0$, so that the excitation
energy $\delta  E$ of the $\bar{n} = 1,2,3,4,\infty$ states explicitly considered here is infinite. This implies that
for {\it all} finite temperatures $T>0$ such that $k_B T < \delta  E = \infty$ the charge 
diffusion constant of the 1D Hubbard model, Eq. (\ref{DproptoOmega}) for $\alpha = \eta$, is finite. Indeed, 
there are no states of class (B) with finite excitation energy to render $\Pi_{\eta} (T)$ infinite at finite $k_B T$
in $D_{\eta} (T) = C_{\eta} (T)\,\Pi_{\eta} (T)$. This is in contrast to the result of hydrodynamic theory and KPZ scaling \cite{Moca_23,Fava_20,Ilievski_18} that $D_{\eta} (T) = \infty$ for $T>0$.

That the energy density $\delta E/L t$ is finite for $z>-1$ is a universal property that applies 
to all states of class (B) of the 1D Hubbard model at the $h = \mu = 0$ point \cite{SM}. 
Our results for finite temperatures $T>0$ are consistent with those of Ref. \cite{Carmelo_24}, which
found that the charge diffusion constant is finite for very small temperatures. The type of charge transport
for larger finite temperatures remained though an important unsolved issue. It was clarified in this Letter
as a result of the identification of the mechanisms that control the type of $T>0$ charge and spin transport.

Our results thus open the door to a key advance in the understanding of the $h=0$ 
spin and $\mu =0$ charge transport in the paradigmatic quantum 
system for low-dimensional strongly correlated electron systems and materials,
such as the 1D Hubbard model.\\

%%%%%%%%%%%%%%%%%%%%%%%%%%%%%%%%%%%%%%%%%%%%%%%%%%%%%%%%%%%%%%%%%%%%%%%%%%
\acknowledgements
We thank Toma\v{z} Prosen for illuminating discussions.
J. M. P. C. and P. D. S. acknowledge support from Funda\c{c}\~ao para a Ci\^encia e Tecnologia
Grant No. UIDB/04650 and Unidade de I{\&}D Grant No.UID/04540 via Centro de F\'{\i}sica e engenharia 
de materiais Avan\c{c}ados, respectively.\\ \\ 
%%%%%%%%%%%%%%%%%%%%%%%%%%%%%%%%%%%%%%%%%%%%%%%%%%%%%%%%%%%%%%%%%%%%%%%%%%

\begin{widetext}

\newpage

\section*{Supplementary material}

The quantities, notations, and units used in this Supplementary Material are those of the Letter.\\ \\

{\bf I - Useful technical/mathematical information needed for the studies of the Letter}\\

Before reporting below in II the information on the $\tau$- and $\alpha$-spin representation \cite{Carmelo_25}
needed for the studies of the Letter, here we provide some useful technical/mathematical information. 
The Bethe-ansatz quantum numbers \cite{Lieb_68,Takahashi_72} 
$I_j^{\tau}$ and $I_j^{\alpha n}$ are the discrete $\tau$-band and $\alpha n$-band momentum values, 
\begin{equation}
q_j = {2\pi\over L} I_j^{\tau}\hspace{0.20cm}{\rm for}\hspace{0.20cm}j = 1,..,L 
\hspace{0.20cm}{\rm and}\hspace{0.20cm}
q_j = {2\pi\over L} I_j^{\alpha n}\hspace{0.20cm}{\rm for}\hspace{0.20cm}j = 1,..,L_{\alpha n} \, ,
\label{qq}
\end{equation}
respectively, in units of $2\pi/L$, where,
\begin{eqnarray}
I_j^{\tau} - {L\over 2\pi} q^{\Delta} & = & 0,\pm 1, \pm 2,...\hspace{0.20cm}{\rm for}\hspace{0.20cm}\sum_{\alpha =s,\eta}M_{\alpha n}\hspace{0.20cm}{\rm even}
\nonumber \\
& = & \mp 1/2,\pm 3/2, \pm 5/2,...\hspace{0.20cm}{\rm for}\hspace{0.20cm}\sum_{\alpha =s,\eta}M_{\alpha n}\hspace{0.20cm}{\rm odd} \, ,
\label{Itau}
\end{eqnarray}
and
\begin{eqnarray}
&& I_j^{\alpha n} = 0,\pm 1, \pm 2,...\hspace{0.20cm}{\rm for}\hspace{0.20cm}2S_{\alpha} + M_{\alpha n}\hspace{0.20cm}{\rm odd}
\nonumber \\
&& \hspace{0.30cm} = \mp 1/2,\pm 3/2, \pm 5/2,...\hspace{0.20cm}{\rm for}\hspace{0.20cm}2S_{\alpha} + M_{\alpha n}\hspace{0.20cm}{\rm even} \, ,
\label{Ian}
\end{eqnarray}
respectively, Here $\alpha = s,\eta$ and $S_s$ is the spin, $S_{\eta}$ the $\eta$-spin, 
$M_{\alpha n}$ the number of $\alpha n$-pairs mentioned in the Letter, $L_{\alpha n} = M_{\alpha n} + M_{\alpha n}^h$,
and $M_{\alpha n}^h$ is the number of $\alpha n$-holes given below in Eq. (\ref{LM}).
The momentum $q^{\Delta}$ appearing in Eq. (\ref{Itau}) is defined below in Eq. (\ref{qqq}). It vanishes in the $u=U/4t\rightarrow 0$
limit and for $u>0$ vanishes in the thermodynamic limit for ground states and excited states generated from them by a 
$\tau$-band finite number of processes.

The number ${\cal{N}}_{\alpha} = L_{\alpha} - 2S_{\alpha}$ of paired physical $\alpha$-spins of an energy eigenstate 
are contained in a number ${\cal{N}}_{\alpha}/2 = L_{\alpha}/2-S_{\alpha}$ of $\alpha$-spin singlet pairs \cite{Carmelo_25}. 
In the thermodynamic limit, such pairs are distributed over $\alpha$-spin singlet configurations, each containing a number
$n=1,....\infty$ of pairs, which we call {\it $\alpha n$-pairs} \cite{Carmelo_25}.
The number of paired physical $\alpha$-spins of projection $\pm 1/2$ can then be written as 
${\cal{N}}_{\alpha,\pm 1/2} = L_{\alpha}/2-S_{\alpha} = \sum_{n=1}^{\infty}n\,M_{\alpha n}$. The values of 
the numbers $M_{\alpha n}$ are for $\alpha = s,\eta$ and $n=1,...,\infty$ specific to each energy eigenstate.

In the thermodynamic limit, the $\tau$-band and $\alpha n$-bands discrete momentum values $q_j$ such that $q_{j+1} - q_j = 2\pi/L$ 
can be treated as a continuous momentum variable $q$. We can then use an alternative rapidity representation $k = k (q)$ for the $\tau$-band and 
$\Lambda = \Lambda_{\alpha n} (q)$ for the $\alpha n$-bands. The continuous rapidity variables
have for all energy eigenstates symmetrical intervals $k \in [-\pi,\pi]$ and $\Lambda\in [-\infty,\infty]$, respectively.
The corresponding $\tau$-band momentum-rapidity-variable distribution ${\bar{N}}_{\tau} (k)$ and $\alpha n$-band 
rapidity-variable distributions ${\bar{M}}_{\alpha n} (\Lambda)$ store the same information
as the $\tau$-band momentum distribution $N_{\tau} (q)$ and 
$\alpha n$-band momentum distributions $M_{\alpha n} (q)$, respectively, considered in the Letter. The former distributions
and the corresponding hole distributions obey the following relations,
\begin{eqnarray}
&& {\bar{N}}_{\tau} (k) = 1 -  {\bar{N}}^h_{\tau} (k)
\hspace{0.20cm} {\rm for}\hspace{0.20cm} k\in [-\pi,\pi]
\hspace{0.20cm} {\rm and}\hspace{0.20cm} 
{\bar{M}}_{\alpha n} (\Lambda) = 1 -  {\bar{M}}^h_{\alpha n} (\Lambda)
\hspace{0.20cm} {\rm for}\hspace{0.20cm} \Lambda\in [-\infty,\infty]
\hspace{0.20cm} {\rm with}
\nonumber \\
&& {\bar{N}}_{\tau} (k (q)) = N_{\tau} (q)\hspace{0.20cm} {\rm and}\hspace{0.20cm}{\bar{M}}_{\alpha n} (\Lambda_{\alpha n} (q)) 
= M_{\alpha n} (q) \hspace{0.20cm}{\rm where}\hspace{0.20cm}\alpha = s,\eta
\hspace{0.20cm} {\rm and}\hspace{0.20cm}n=1,...,\infty \, .
\label{NNrela}
\end{eqnarray}

The $\tau$-band momentum rapidity function $k (q)$ and $\alpha n$-bands rapidity functions $\Lambda_{\alpha n} (q)$
appearing here can be defined by their inverse functions $q_{\tau} (k)$ and $q_{\alpha n} (\Lambda)$, respectively. 
For the $\tau$-, $sn$-, and $\eta n$-bands the latter functions are defined 
by coupled Bethe-ansatz equations expressed in functional form in Eqs. (A5)-(A7) of Ref. \onlinecite{Carmelo_25}.
The functions $2\pi\rho (k) = {d q_{\tau} (k)\over dk}$ and $\sigma_{\alpha n}  (\Lambda) =  {d q_{\alpha n} (\Lambda)\over d\Lambda}$ 
appearing in these equations are Jacobians associated with the transformation from $\tau$-band momentum $q$ and $\alpha n$-bands momentum $q$ 
to the rapidity variables $k$ and $\Lambda$, respectively. Such functions are defined by coupled integral equations,
Eqs. (A9)-(A11) of Ref. \onlinecite{Carmelo_25}.

The $\tau$-band $j=1,...,L$ and $\alpha n$-bands $j=1,...,L_{\alpha n}$ discrete set of $q_j$'s
such that $q_{j+1}-q_j = 2\pi/L$, Eqs. (\ref{qq})-(\ref{Ian}), have values in the intervals 
$q_j \in [q_{\tau}^-,q_{\tau}^+]$ and $q_j \in [q_{\alpha n}^-,q_{\alpha n}^+]$, 
respectively, where,
\begin{eqnarray}
q_{\tau}^{\pm} & = & q_{\tau 0}^{\pm}  + q^{\Delta} 
\hspace{0.20cm}{\rm with}\hspace{0.20cm}q^{\Delta}  = q_{\tau} (0)
\nonumber \\
q_{\tau 0}^{\pm} & = & \pm{\pi\over L} (L-1) 
\hspace{0.20cm}{\rm for}\hspace{0.20cm}\sum_{\alpha =s,\eta}\sum_{n=1}^{\infty}M_{\alpha n}\hspace{0.20cm}{\rm odd}
\nonumber \\
q_{\tau 0}^+ & = & + \pi \hspace{0.20cm}{\rm and}\hspace{0.20cm}q_{\tau 0}^- = - {\pi\over L} (L - 2)
\hspace{0.20cm}{\rm for}\sum_{\alpha =s,\eta}\sum_{n=1}^{\infty}M_{\alpha n}\hspace{0.20cm}{\rm even}
\nonumber \\
q_{\alpha n}^{\pm} & = & \pm {\pi\over L}(L_{\alpha n} -1) 
\hspace{0.20cm}{\rm for}\hspace{0.20cm}\alpha = s,\eta\hspace{0.20cm}{\rm and}\hspace{0.20cm}n = 1,....\infty \, .
\label{qqq}
\end{eqnarray}
In the thermodynamic limit, the $\tau$-band interval simplifies to $q_j \in [-\pi + q^{\Delta},\pi + q^{\Delta}]$ and those
of the $\alpha n$-bands to $q_j \in [-\pi L_{\alpha n}/L, \pi L_{\alpha n}/L]$.
The $\tau$-band momentum $q^{\Delta}$ reads $q^{\Delta} = q_{\tau} (0)$ where $q_{\tau} (k)$ is defined by
Eq. (A5) of Ref. \onlinecite{Carmelo_25}. As mentioned above, for ground states and excited states generated 
from them by a $\tau$-band finite number of processes, $q^{\Delta}=0$ in the thermodynamic limit.

The relation between the limiting values of the rapidity variables and
those of the momentum values $q_j$, Eqs. (\ref{qq})-(\ref{Ian}), reads,
\begin{equation} 
k (q_{\tau}^{\pm}) = \pm\pi \hspace{0.20cm}{\rm and}\hspace{0.20cm} 
\Lambda_{\alpha n} (q_{\alpha n}^{\pm}) = \pm\infty \, .
\label{kqLq}
\end{equation}

The internal degrees of freedom of one $\alpha n$-pair with (i) $n=1$ and (ii) $n>1$ $\alpha$-spin singlet pairs of
energy eigenstates for which $M_{\alpha n}>0$ correspond to (i) one unbound $\alpha$-spin singlet pair and 
(ii) a number $l = 1,...,n$ of bound $\alpha$-spin singlet pairs, respectively. In the thermodynamic limit, the structure of
the complex Bethe-ansatz rapidities simplifies. They are given by \cite{Takahashi_72},
\begin{equation}
\Lambda_{\alpha n,l} (q) = \Lambda_{\alpha n} (q) + i(n + 1 -2l)\,u \hspace{0.20cm}{\rm where}\hspace{0.20cm}l=1,...,n \, .
\label{LambdaIm}
\end{equation}
Here the index $l=1,...,n$ labels each of the $n=1,...,\infty$ $\alpha$-spin singlet pairs contained in one $\alpha n$-pair.
The rapidity $\Lambda_{\alpha n,l} (q)$ is real for $n=1$ and for $n>1$ its imaginary part describes the binding of
the corresponding $l=1,...,n$ $\alpha$-spin singlet pairs. Its real part is the rapidity function $\Lambda_{\alpha n} (q)$ the inverse function 
of which $q_{\alpha n} (\Lambda)$ is defined by the Bethe-ansatz equations, 
Eqs. (A6) and (A7) of Ref. \onlinecite{Carmelo_25} for $\alpha =s$ and  $\alpha =\eta$, respectively.\\ 

{\bf II - The physical $\alpha$-spins, the model's global $[SU (2)\times SU(2)\times U(1)]/Z_2^2$ symmery, and the
$\tau$- and $\alpha$-spin representation}\\

In the presence of a magnetic field $h$ and chemical potential $\mu$, the Hamiltonian of the 1D Hubbard model provided in Eq. (2) 
of the Letter for $h=\mu=0$ is under periodic boundary conditions given by,
\begin{equation}
\hat{H} = -t\sum_{\sigma, j}\left[c_{j,\sigma}^{\dag}\,c_{j+1,\sigma} + 
{\rm h.c.}\right] + U\sum_{j}\hat{\rho}_{j,\uparrow}\hat{\rho}_{j,\downarrow} -
\sum_{\alpha = s,\eta}\mu_{\alpha} {\hat{S}}_{\alpha}^{z} \, .
\label{H}
\end{equation}
Here $\alpha = s$ refers to spin and $\alpha = \eta$ to $\eta$-spin,
$\mu_{s} = 2\mu_B h$ where $\mu_B$ is the Bohr magneton,
$\mu_{\eta} = 2\mu$, $c_{j,\sigma}^{\dag}$ creates one electron of spin projection $\sigma$ at site $j$,
$\hat{\rho}_{j,\sigma}= (\hat{n}_{j,\sigma}-1/2)$, and $\hat{n}_{j,\sigma}=c_{j,\sigma}^{\dag}\,c_{j,\sigma}$.

We can generate from any of the $4^L$ $u>0$ energy eigenstates $\vert\Psi,u\rangle$ of the 1D Hubbard model, Eq. (\ref{H}), 
a corresponding energy eigenstate $\vert\Psi,\infty\rangle$ for $u\rightarrow\infty$, 
as $\vert\Psi,\infty\rangle = {\hat{V}}_u\vert\Psi,u\rangle$ \cite{Carmelo_25}. The $u$-unitary operator ${\hat{V}}_u$ is
uniquely defined in Eq. (11) of Ref. \onlinecite{Carmelo_17A} by its matrix elements between the model's $4^L$ 
finite-$u$ energy and momentum eigenstates. The Bethe ansatz performs the $u$-unitary transformation,
consistent with such matrix elements involving that ansatz quantities.
There are many choices for sets of $4^L$ energy eigenstates for $u\rightarrow\infty$. The $\alpha$-spin representation refers to
that defined here. Reciprocally, any $u>0$ energy eigenstates $\vert\Psi,u\rangle$ can be generated as 
$\vert\Psi,u\rangle = {\hat{V}}_u^{\dag}\vert\Psi,\infty\rangle$. 

The $u>0$ (i) number $L_{s,\pm 1/2}$ of physical spins of projection $\pm 1/2$ and (ii) $L_{\eta,\pm 1/2}$ 
%CORRPOSPUBLY "of" added
of $\eta$-spins of projection $\pm 1/2$ 
are generated through $\vert\Psi,u\rangle = {\hat{V}}_u^{\dag}\vert\Psi,\infty\rangle$ from the $u\rightarrow\infty$  
(i) number $L_s/2 \pm S_s^z$ of sites singly occupied by electrons of spin projection $\pm 1/2$ and (ii) 
$L_{\eta}/2 \pm S_{\eta}^z$ of empty sites associated with $\eta$-spin-projection $+1/2$ and sites doubly occupied 
by electrons associated with $\eta$-spin-projection $-1/2$, 
respectively. Hence, the numbers $L_s$ of physical spins and $L_{\eta}$ of physical $\eta$-spins such that $L_s + L_{\eta} = L$
are also generated through $\vert\Psi,u\rangle = {\hat{V}}_u^{\dag}\vert\Psi,\infty\rangle$ under the relations
$L_s = L_{s,+1/2}+ L_{s,-1/2}$ and $L_{\eta} = L_{\eta,+ 1/2} + L_{\eta, -1/2}$, respectively.

For $u>0$, the $u$-independent quantum numbers $q_j = {2\pi\over L} I_j^{\tau}$ and $q_j = {2\pi\over L} I_j^{\alpha n}$,
Eq. (\ref{qq}), the occupancy configurations of which generate the energy eigenstates $\vert\Psi,u\rangle$,
stem from corresponding $u\rightarrow\infty$ energy eigenstates $\vert\Psi,\infty\rangle$. The former $u>0$ states 
$\vert\Psi,u\rangle = {\hat{V}}_u^{\dag}\vert\Psi,\infty\rangle$ have exactly the same values and occupancy configurations 
of the $u$-independent quantum numbers $q_j = {2\pi\over L} I_j^{\tau}$ and $q_j = {2\pi\over L} I_j^{\alpha n}$
as the $u\rightarrow\infty$ energy eigenstate from which they have been generated through the unitary
operator ${\hat{V}}_u^{\dag}$.

The $\tau$- and $\alpha$-spin representation \cite{Carmelo_25} used in the studies of the Letter in terms of such $L_s = L_{s,+1/2}+ L_{s,-1/2}$ 
physical spins and $L_{\eta} = L_{\eta,+ 1/2} + L_{\eta, -1/2}$ physical $\eta$-spins refers to the whole Hilbert space of
the 1D Hubbard model. On the other hand, the Bethe ansatz \cite{Lieb_68,Takahashi_72}
refers only to subspaces spanned by either the $\alpha$-spin lowest-weight 
states (LWSs) or the $\alpha$-spin  highest-weight states (HWSs) of the two, $\alpha =s,\eta$, $\alpha$-spin $SU(2)$ algebras. 

It is convenient to denote the $u>0$ energy and momentum eigenstates $\vert\Psi,u\rangle$
by $\vert l_{\rm r}^u,S_{\tau},S_{\alpha},S_{\alpha}^z,S_{\bar{\alpha}},S_{\bar{\alpha}}^z\rangle$.
Here $\bar{\eta} = s$ and $\bar{s} = \eta$, the eigenvalue of the $\tau$-translational $U(1)$ symmetry's generator reads 
$S_{\tau} = {1\over 2}L_{\eta} = {1\over 2}(L - L_s)$, and $l_{\rm r}^u$ stands for $u=U/4t>0$ and all 
$u$-independent quantum numbers other than $S_{\tau},S_{\alpha},S_{\alpha}^z,S_{\bar{\alpha}}$, and $S_{\bar{\alpha}}^z$
needed to uniquely specify each energy and momentum eigenstate.

Any $\alpha$-spin non-HWS $\vert l_{\rm r}^u,S_{\tau},S_{\alpha},S_{\alpha}^z,S_{\bar{\alpha}},S_{\bar{\alpha}}^z\rangle$
can be generated from a corresponding HWS $\vert l_{\rm r}^u,S_{\tau},S_{\alpha},S_{\alpha},S_{\bar{\alpha}},S_{\bar{\alpha}}^z\rangle$ 
as follows,
\begin{eqnarray}
&& \vert l_{\rm r}^u,S_{\tau},S_{\alpha},S_{\alpha}^z,S_{\bar{\alpha}},S_{\bar{\alpha}}^z\rangle = 
\left[\frac{1}{\sqrt{{\cal{C}}_{\alpha}}}({\hat{S}}^{+}_{\alpha})^{n_{\alpha}^z}\right]
\vert l_{\rm r}^u,S_{\tau},S_{\alpha},S_{\alpha},S_{\bar{\alpha}},S_{\bar{\alpha}}^z\rangle 
\nonumber \\
&& {\rm where}\hspace{0.20cm}\bar{\eta} = s \hspace{0.20cm}{\rm and}\hspace{0.20cm}\bar{s} = \eta \, .
\label{Gstate-BAstate}
\end{eqnarray}
Here the normalization constant reads ${\cal{C}}_{\alpha} = [n_{\alpha}^z!]\prod_{j=1}^{n_{\alpha}^z}[\,2S_{\alpha}+1-j\,]$,
${\hat{S}}_{\alpha}^{+}$ is the $\alpha$-spin $SU(2)$ off-diagonal generator of $\alpha$-spin flips, and
$n_{\alpha}^z = S_{\alpha} - S_{\alpha}^z = 0,1,..., 2S_{\alpha} $.

While when $\mu = h =0$ the Hamiltonian, Eq. (\ref{H}), commutes with the generators of the 
global $[SU (2)\times SU(2)\times U(1)]/Z_2^2$ symmetry \cite{Carmelo_25,Carmelo_10}, 
the $4^L$ irreducible representations of that symmetry refer to a
complete set of $4^L$ energy eigenstates for all values of the chemical potential $\mu$ and magnetic
field $h$. Therefore, the dimension of the full Hilbert space of the Hamiltonian, Eq. (\ref{H}), is obtained by the summation over the
integer values of $2S_{\tau}\geq 0$, $2S_s\geq 0$, and $2S_{\eta}\geq 0$ of the product of the numbers of irreducible
representations of each of the three global symmetries in $[SU (2)\times SU(2)\times U(1)]/Z_2^2$, as given in Eqs. (A1)-(A4)
of Ref. \onlinecite{Carmelo_25}.\\

{\bf III - The $\alpha$-spin carriers and the general expressions for their $\alpha$-spin elementary currents}\\

Here we summarize some of the information on $\alpha$-spin carriers and
$\alpha$-spin elementary currents provided in Ref. \onlinecite{Carmelo_25} needed for the studies of the Letter.

The Hamiltonian, Eq. (\ref{H}), in the presence of a uniform vector potential (twisted boundary conditions) remains solvable 
by the Bethe ansatz \cite{Shastry_90,Carmelo_18}. In the case of coupling to the vector potential associated with spin, 
$\Phi = \Phi_{\uparrow}=\Phi_{\downarrow}$,we then find  the following deviations in the presence of 
that potential: $q_j \rightarrow q_j + {\Phi\over L}$ for the $\tau$-band, $q_j \rightarrow q_j - {2n\Phi\over L}$ for 
the $sn$-bands, and $q_j \rightarrow q_j$ for the $\eta n$-bands. For coupling to the vector potential associated with 
$\eta$-spin/charge, $\Phi = \Phi_{\uparrow}=-\Phi_{\downarrow}$,  
the deviations read $q_j \rightarrow q_j + {\Phi\over L}$ for the $\tau$-band, $q_j \rightarrow q_j$ for the $sn$-bands, and 
$q_j \rightarrow q_j - {2n\Phi\over L}$ for the $\eta n$-bands. Importantly, the $\tau$-band quantum numbers 
$q_j$ couple {\it both} to the vector potentials associated with spin and charge/$\eta$-spin.

Consistent, we find that the deviation $P_{\Phi} - P$ where $P_{\Phi}$ and $P$ are the momentum eigenvalues
for $\Phi/L\neq 0$ and $\Phi/L= 0$, respectively, is for HWSs given by,
\begin{equation}
P_{\Phi} - P  = \left(L_{\alpha}-\sum_{n=1}^{\infty}2n\,M_{\alpha n}\right)\iota_{\alpha}{\Phi\over L}
\hspace{0.20cm}{\rm for}\hspace{0.20cm}\alpha = s,\eta\hspace{0.20cm}{\rm where}\hspace{0.20cm}\
\iota_{s} = 1\hspace{0.20cm}{\rm and}\hspace{0.20cm}\iota_{\eta} = -1 \, .
\label{PeffUalpha}
\end{equation}
Here the number multiplying $\iota_{\alpha}\,\Phi/L$ is that of physical $\alpha$-spins that couple to the vector potential
associated with $\alpha$-spin.

The term $L_{\alpha}\,\iota_{\alpha}{\Phi\over L}$ in $(L_{\alpha}-\sum_{n}2n\,M_{\alpha n})\iota_{\alpha}{\Phi\over L}$
refers to {\it all} $L_{\alpha}$ physical $\alpha$-spins coupling to the vector potential in the absence of $\alpha$-spin singlet pairing. 
On the other hand, the coupling counter terms $-\sum_{n}2n\,M_{\alpha n}\iota_{\alpha}{\Phi\over L}$ 
refer to the number $2n$ of paired physical $\alpha$-spins in each $\alpha n$-pair. 
They {\it exactly cancel} the coupling in $L_{\alpha}\,\iota_{\alpha}{\Phi\over L}$ of the 
corresponding number $2n$ of paired $\alpha$-spins in each such an $\alpha n$-pair. 
The exact relation, $2S_{\alpha} = L_{\alpha}-\sum_{n=1}^{\infty}2n\,M_{\alpha n}$,
then reveals that {\it only} the number $N_{\alpha} = 2S_{\alpha}$ of unpaired physical $\alpha$-spins 
of any $S_{\alpha}>0$ energy eigenstate couple to the vector potential associated with $\alpha$-spin. 
Hence they are the {\it $\alpha$-spin carriers}.

In the case of $\alpha$-spin non-HWSs, the use of Eq. (\ref{Gstate-BAstate}) leads to the following more general expression
for $(P_{\Phi} - P)$,
\begin{equation}
P_{\Phi} - P = (N_{\alpha,+1/2}-N_{\alpha,-1/2})\,\iota_{\alpha}{\Phi\over L}
\hspace{0.20cm}{\rm for}\hspace{0.20cm}\alpha = s,\eta \, .
\label{PeffUalphaNnonHWS}
\end{equation}
This shows that the coupling to the vector potential associated with $\alpha$-spin of the $\alpha$-spin carriers with  opposite projection $\pm 1/2$ 
has opposite sign.

The $z$ component of the spin and charge current operators are in units of spin $1/2$ and electronic
charge given by,
\begin{equation}
\hat{J}_s^z = - it\sum_{\sigma}\sum_{j=1}^L(2\sigma)\left[c_{j,\sigma}^{\dag}\,c_{j+1,\sigma} -
c_{j+1,\sigma}^{\dag}\,c_{j,\sigma}\right]
\hspace{0.20cm}{\rm and}\hspace{0.20cm}
\hat{J}_{\eta}^z = - it\sum_{\sigma}\sum_{j=1}^L\left[c_{j,\sigma}^{\dag}\,c_{j+1,\sigma} -
c_{j+1,\sigma}^{\dag}\,c_{j,\sigma}\right] \, ,
\label{Jzseta}
\end{equation}
respectively. In the factor $(2\sigma)$ of the spin current operator expression, $\sigma$ reads 
$+1/2$ for $\sigma = \uparrow$ and $-1/2$ for $\sigma = \downarrow$.

The current operators, Eq. (\ref{Jzseta}), do not commute with the 1D Hubbard model Hamiltonian, Eq. (\ref{H}).
To derive the general expressions of the zero-temperature $\alpha$-spin current expectation values $\langle \hat{J}_{\alpha}^z\rangle$
for both HWSs and non-HWSs, we have first calculated those of HWSs that we denote by $\langle\hat{J}_{\alpha}^z(HWS)\rangle$. The latter 
are obtained by taking the $\Phi/L \rightarrow 0$ limit of the derivative of the HWSs energy eigenvalues multiplied by $-1$ with respect to $\Phi/L$
of the Hamiltonian, Eq. (\ref{H}), in the presence of a uniform vector potential (twisted boundary conditions).
Next, we combined the obtained general expression for $\langle\hat{J}_{\alpha}^z(HWS)\rangle$
with the generation of non-HWSs from the HWSs given in Eq. (\ref{Gstate-BAstate}). We then
arrived to the following general expressions for the current expectation values $\langle \hat{J}_{s}^z\rangle$
and $\langle \hat{J}_{\eta}^z\rangle$ \cite{Carmelo_25},
%CORRPOSPUBLY signs corrected
\begin{equation}
 \langle \hat{J}_{s}^z\rangle = 2t\, c_{\tau}^{n_s^z}\sum_{j=1}^L {N_{\tau} (q_j)\over 2\pi\rho (k_j)} \sin k_j
+ {(4t)^2\over U}{2\over L}\sum_{n=1}^{\infty} c_{sn}^{n_s^z}
 \sum_{j=1}^{L_{sn}} {M_{sn}^h (q_j)\over 2\pi\sigma_{sn} (\Lambda_{sn,j})}\sum_{j'=1}^L
 {N_{\tau} (q_{j'}) \sin k_{j'}\over 1 + \left({\Lambda_{sn,j} - \sin k_{j'}\over nu}\right)^2} \, ,
 \label{Jszq}
 \end{equation}
 and
 \begin{eqnarray}
 \langle \hat{J}_{\eta}^z\rangle & = & - 2t\, c_{\tau}^{n_{\eta}^z}\sum_{j=1}^L {N_{\tau}^h (q_j)\over 2\pi\rho (k_j)} \sin k_j
 + {8t\over L}\sum_{n=1}^{\infty}c_{\eta n}^{n_{\eta}^z} \sum_{j=1}^{L_{\eta n}} {M_{\eta n}^h (q_j)\over
 2\pi\sigma_{\eta n}  (\Lambda_{\eta n,j})}
 \nonumber \\
 & \times & \Bigl(n\,{\rm Re}\Bigl\{{(\Lambda_{\eta n,j} - i n u)\over\sqrt{1 - (\Lambda_{\eta n,j} - i n u)^2}}\Bigr\}
 + {4t\over U} \sum_{j'=1}^L {N_{\tau} (q_{j'}) \sin k_{j'}\over 1 + \left({\Lambda_{\eta n,j} - \sin k_{j'}\over nu}\right)^2}\Bigr) \, ,
 \label{Jetazq}
 \end{eqnarray}
respectively. Here $k_j = k (q_j)$, $\Lambda_{\eta n,j} = \Lambda_{\eta n} (q_j)$, 
\begin{equation}
c_{\tau}^{n_{\alpha}^z} = {2S_{\alpha}^z\over L_{\alpha}}
\hspace{0.20cm}{\rm and}\hspace{0.20cm}
c_{\alpha n}^{n_{\alpha}^z} = {2S_{\alpha}^z\over M_{\alpha n}^h} 
\hspace{0.20cm}{\rm where}\hspace{0.20cm}n_{\alpha}^z = S_{\alpha} - S_{\alpha}^z
\hspace{0.20cm}{\rm and}\hspace{0.20cm}2S_{\alpha}^z = N_{\alpha,+1/2} - N_{\alpha,-1/2} \, ,
\label{cc}
\end{equation}
where $c_{\tau}^{n_{\alpha}^z}$ and $c_{\alpha n}^{n_{\alpha}^z}$ are coupling factors of the $\alpha$-spin carriers, as justified below.

The general expressions for the HWS's $\alpha$-spin current expectation values 
$\langle\hat{J}_{\alpha}^z(HWS)\rangle$ have exactly the same form as those given in Eqs. (\ref{Jszq}) and (\ref{Jetazq})
with $n_{s}^z =0$ and $n_{\eta}^z =0$, respectively, in the coupling factors, Eq. (\ref{cc}). For the HWSs
these factors then read $c_{\tau}^{0} = {2S_{\alpha}\over L_{\alpha}}$
and $c_{\alpha n}^{0} = {2S_{\alpha}\over M_{\alpha n}^h}$ for $\alpha = s,\eta$. 
Indeed, for a HWS all $N_{\alpha} = 2S_{\alpha}$ $\alpha$-spin carriers have the same projection $+1/2$.
The translational degrees of freedom of such $\alpha$-spin carriers
are described both by a number $L_{\alpha}$ of $\tau$-band momentum values $q_j$
and a number $M_{\alpha n}^h$ of $\alpha n$-holes momentum values $q_j$ for each $\alpha n$-band
for which $M_{\alpha n} >0$. These numbers read,
\begin{equation}
L_{\alpha} = 2S_{\alpha} + \sum_{n=1}^{\infty}2n\,M_{\alpha n}\hspace{0.20cm}{\rm and}\hspace{0.20cm}
M_{\alpha n}^h = 2S_{\alpha} + \sum_{n'=n+1}^{\infty}2(n' - n)\,M_{\alpha n'} \, .
\label{LM}
\end{equation}

The contributions to the HWSs $\alpha$-spin current expectation values $\langle\hat{J}_{\alpha}^z(HWS)\rangle$
stem from relative configurations of the $\tau$-band $j = 1,...,L$ momentum values $q_j$ associated with
the $L_{\alpha}$ physical $\alpha$-spins and $L_{\bar{\alpha}}$ physical $\bar{\alpha}$-spins, respectively, where $\bar{\alpha}$ is defined 
in Eq. (\ref{Gstate-BAstate}). However, it follows from Eq. (\ref{PeffUalpha}) that only a number $N_{\alpha} = 2S_{\alpha} < L_{\alpha}$ of unpaired 
physical $\alpha$-spins on the right-hand side of $L_{\alpha}$'s expression in Eq. (\ref{LM}) out of the total number $L_{\alpha}$ 
of physical $\alpha$-spins couple to the vector potential associated with $\alpha$-spin and thus contribute to such expectation values. 

The HWSs $\alpha$-spin current expectation values $\langle\hat{J}_{\alpha}^z(HWS)\rangle$ also have contributions 
from $\alpha n$-bands  for which $M_{\alpha n} >0$ with $j = 1,...,L_{\alpha n}$ momentum values $q_j$. The latter contributions 
stem from relative configurations of such $j = 1,...,L_{\alpha n}$ momentum values $q_j$ associated with $M_{\alpha n}^h$ 
$\alpha n$-holes and $M_{\alpha n}$ $\alpha n$-pairs, respectively.
Again, only a number $N_{\alpha} = 2S_{\alpha} < M_{\alpha n}^h$ of such $M_{\alpha n}^h$ $\alpha n$-holes on the right-hand side of 
the $M_{\alpha n}^h$'s expression in Eq. (\ref{LM}) describe the degrees of freedom of the number $N_{\alpha} = 2S_{\alpha} < M_{\alpha n}^h$ 
of unpaired physical $\alpha$-spins in $S_{\alpha}>0$ energy eigenstates that couple to the vector potential associated with $\alpha$-spin.

The point is that there is quantum uncertainty concerning which of the number $L_{\alpha}$ of $\tau$-band momentum values $q_j$
associated with all $L_{\alpha}$ physical $\alpha$-spins in Eq. (\ref{LM})
and which of the number $M_{\alpha n}^h$ of $\alpha n$-band momentum values $q_j$ associated with $\alpha n$-holes in that equation
describe the number $N_{\alpha} = 2S_{\alpha}$ of unpaired physical $\alpha$-spins that couple to 
the vector potential associated with $\alpha$-spin.

That only they couple to the vector potential associated with $\alpha$-spin and thus contribute 
to the HWSs current expectation values $\langle\hat{J}_{\alpha}^z(HWS)\rangle$
is then ensured by quantum effects within the Bethe ansatz. Consistent with
Eq. (\ref{PeffUalpha}), such effects lead to the coupling factors $c_{\tau}^{0} = {2S_{\alpha}\over L_{\alpha}}$ and $c_{\alpha n}^{0} = 
{2S_{\alpha}\over M_{\alpha n}^h}$, respectively, in such current expectation values expression. In the case of non-HWSs,
there are unpaired physical $\alpha$-spins with opposite projections $\pm 1/2$ the couplings of which to 
the vector potential associated with $\alpha$-spin
have opposite sign. Consistent with Eq. (\ref{PeffUalphaNnonHWS}), combination of the quantum effects within the Bethe ansatz
with the $\alpha$-spin flip processes, Eq. (\ref{Gstate-BAstate}), leads to the replacement of 
$2S_{\alpha} = N_{\alpha}$ to $2S_{\alpha}^z = N_{\alpha,+1/2} - N_{\alpha,-1/2}$ in the coupling factors
given in Eq. (\ref{cc}).

The non-HWSs current expectation values, Eqs. (\ref{Jszq}) and (\ref{Jetazq}),
can be written as $\langle\hat{J}_{\alpha}^z\rangle = {S_{\alpha}^z\over S_{\alpha}}\langle\hat{J}_{\alpha}^z(HWS)\rangle$
or equivalently as $\langle\hat{J}_{\alpha}^z\rangle = \sum_{\sigma =\pm 1/2} j_{\alpha,\sigma}\,N_{\alpha,\sigma}$
where $j_{\alpha,\sigma}$ is given by \cite{Carmelo_25},
\begin{equation}
j_{\alpha,\pm 1/2}
 = \pm {\langle\hat{J}_{\alpha}^z(HWS)\rangle\over N_{\alpha}} =  = \pm {\langle\hat{J}_{\alpha}^z(HWS)\rangle\over 2S_{\alpha}} \, .
\label{J-eta-spin-spin}
\end{equation}
It is then straightforward to confirm that $j_{\alpha,\pm 1/2}$ 
is the $\alpha$-spin elementary current carried by one unpaired physical 
$\alpha$-spin of projection $\pm 1/2$. Indeed, under each unpaired physical $\alpha$-spin flip generated by the operator 
${\hat{S}}_{\alpha}^{+}$ in Eq. (\ref{Gstate-BAstate}), the $\alpha$-spin current expectation value
$\langle\hat{J}_{\alpha}^z\rangle = {S_{\alpha}^z\over S_{\alpha}}\langle\hat{J}_{\alpha}^z(HWS)\rangle$
exactly changes by $-2j_{\alpha,+1/2} = 2j_{\alpha,-1/2}$.

The general expression of the $\alpha$-spin elementary currents carried by the $\alpha$-spin carriers
is straightforwardly obtained by dividing the general expression of $\pm\langle\hat{J}_{\alpha}^z(HWS)\rangle$ 
by the number $N_{\alpha} = 2S_{\alpha}$ of $\alpha$-spin carriers,
as given in Eq. (\ref{J-eta-spin-spin}). For $\alpha =s$ and $\alpha =\eta $ this gives \cite{Carmelo_25},
\begin{equation}
 j_{s,\pm 1/2} = \pm {2t\over N_{\tau}}\sum_{j=1}^L {N_{\tau} (q_j)\over 2\pi\rho (k_j)} \sin k_j
\pm {(4t)^2\over U}{2\over L}\sum_{n=1}^{\infty} {1\over M_{sn}^h}
 \sum_{j=1}^{L_{sn}} {M_{sn}^h (q_j)\over 2\pi\sigma_{sn} (\Lambda_{sn,j})}\sum_{j'=1}^L
 {N_{\tau} (q_{j'}) \sin k_{j'}\over 1 + \left({\Lambda_{sn,j} - \sin k_{j'}\over nu}\right)^2} \, ,
 \label{jsqj}
 \end{equation}
 and
 \begin{eqnarray}
 j_{\eta,\pm 1/2} & = & \mp {2t\over N_{\tau}^h}\sum_{j=1}^L {N_{\tau}^h (q_j)\over 2\pi\rho (k_j)} \sin k_j
 \pm {8t\over L}\sum_{n=1}^{\infty} {1\over M_{\eta n}^h}
 \sum_{j=1}^{L_{\eta n}} {M_{\eta n}^h (q_j)\over 2\pi\sigma_{\eta n}  (\Lambda_{\eta n,j})}
 \nonumber \\
 & \times & \Bigl(n\,{\rm Re}\Bigl\{{(\Lambda_{\eta n,j} - i n u)\over\sqrt{1 - (\Lambda_{\eta n,j} - i n u)^2}}\Bigr\}
 + {4t\over U} \sum_{j'=1}^L {N_{\tau} (q_{j'}) \sin k_{j'}\over 1 + \left({\Lambda_{\eta n,j} - \sin k_{j'}\over nu}\right)^2}\Bigr) \, ,
 \label{jetaqj}
 \end{eqnarray}
respectively. 

The momentum distributions $N_{\tau} (q_j)$, $M_{s n}^h (q_j)$, $N_{\tau}^h (q_j)$, and 
$M_{\eta n}^h (q_j)$ in these expressions describe all occupancy configurations of the 
discrete $\tau$-band and $\alpha =s,\eta$ $\alpha n$-band momentum values $q_j$ that generate different energy eigenstates.

As mentioned in the Letter, most $\alpha$-spin elementary currents, Eqs. (\ref{jsqj}) and (\ref{jetaqj}), are of the order of $1/L$ and
vanish in the thermodynamic limit. This is because most $\alpha n$-bands occupancy
configurations that generate the energy eigenstates have both occupied and unoccupied discrete momentum
values for $q_j <0$ and $q_j >0$. This implies that many of the contributions to the $\alpha$-spin elementary currents
from $q_j <0$ and $q_j >0$ occupancies cancel each other. In contrast, the states of class (B) have compact occupancies 
of $\alpha n$-holes and compact configurations of $\alpha n$-pairs with momentum values $q_j$ of opposite
sign that provide finite contributions to the $\alpha$-spin elementary currents. 

The states of class (B) with largest finite $\alpha$-spin elementary currents have in rapidity space 
$\alpha n$-bands the $\alpha n$-holes of which have compact occupancy for $\Lambda <0$ and the $\alpha n$-pairs
of which have compact occupancy for $\Lambda >0$, respectively, or vice versa.  
This is the case of the $\bar{n}$ states of class (B) considered in the Letter. It is then useful to express the 
$\alpha$-spin elementary currents, Eqs. (\ref{jsqj}) and (\ref{jetaqj}), in terms of the 
rapidity representation, Eq. (\ref{NNrela}). In the thermodynamic limit, this gives,
\begin{equation}
 j_{s,\pm 1/2} = \pm {t\over\pi}{L\over N_{\tau}}\int_{-\pi}^{\pi}dk {\bar{N}}_{\tau} (k) \sin k
\pm {2t\over \pi^2 u}\sum_{n=1}^{\infty} {L\over M_{sn}^h}
 \int_{-\infty}^{\infty}d\Lambda {\bar{M}}_{sn}^h (\Lambda)\int_{-\pi}^{\pi}dk 
 {\bar{N}}_{\tau} (k) {2\pi\rho (k) \sin k\over 1 + \left({\Lambda - \sin k\over nu}\right)^2} \, ,
 \label{js}
 \end{equation}
 and
 \begin{eqnarray}
j_{\eta,\pm 1/2} & = & \mp {t\over\pi}{L\over N_{\tau}^h}\int_{-\pi}^{\pi}dk {\bar{N}}_{\tau}^h (k) \sin k
 \pm {4t\over \pi}\sum_{n=1}^{\infty} {L\over M_{\eta n}^h}\int_{-\infty}^{\infty}d\Lambda
 {\bar{M}}_{\eta n}^h (\Lambda)
 \nonumber \\
 && \times \Bigl(n\,{\rm Re}\Bigl\{{(\Lambda - i n u)\over\sqrt{1 - (\Lambda - i n u)^2}}\Bigr\}
+ {1\over 2\pi u} \int_{-\pi}^{\pi}dk {\bar{N}}_{\tau} (k) {2\pi\rho (k) \sin k\over 1 + \left({\Lambda - \sin k\over nu}\right)^2}\Bigr) \, ,
 \label{jeta}
 \end{eqnarray}
respectively.\\

{\bf IV - The $\alpha$-spin elementary currents of the $\bar{n}$ states of classes (B) considered in the Letter}\\

The $\bar{n}$ states momentum-rapidity and rapidity-variable distributions given in Eq. (5) of the Letter 
correspond to momentum distributions given by,
\begin{eqnarray}
N_{\tau} (q) & = & 1 \hspace{0.20cm}{\rm for}\hspace{0.20cm}q - q^{\Delta} \in [\pi z,\pi]
\nonumber \\
& = & 0\hspace{0.20cm}{\rm for}\hspace{0.20cm}q - q^{\Delta}\in [-\pi,\pi z]
\nonumber \\
M_{\alpha n} (q) & = & 1 \hspace{0.20cm}{\rm for}\hspace{0.20cm}q\in [q_{\alpha n}^0,q_{\alpha n}]
\hspace{0.20cm}{\rm and}\hspace{0.20cm}n = 1,...,\bar{n}
\nonumber \\
& = & 0\hspace{0.20cm}{\rm for}\hspace{0.20cm}
q\in [-q_{\alpha n},q_{\alpha n}^0]\hspace{0.20cm}{\rm and}\hspace{0.20cm}n = 1,...,\bar{n}
\nonumber \\
M_{\alpha n} (q) & = & 0\hspace{0.20cm}{\rm for}\hspace{0.20cm}n > \bar{n} \, ,
\label{NNOstates1}
\end{eqnarray}
both for $\alpha =s$ and $\alpha =\eta$ where the $\tau$-band shifted momentum $q - q^{\Delta} = \pi z$ 
can have values corresponding to $z = (L_{\eta} - L_s)/L \in [-1,1]$. As reported in the Letter, there is a related variable
$y = y (z) \in [-1,1]$ associated with the $\tau$-band momentum rapidity function $k (q)$ given by 
$y = y (z) = k (\pi z + q^{\Delta}) \in [-1,1]$. The $\alpha n$-band separate momentum $q_{\alpha n}^0$ 
in Eq. (\ref{NNOstates1}) is such that the $\alpha n$-band rapidity function $\Lambda_{\alpha n} (q)$
vanishes at $q - q^{\Delta} = q_{\alpha n}^0$, $\Lambda_{\alpha n} (q_{\alpha n}^0 + q^{\Delta}) = 0$.
It reads $q_{\alpha n}^0= \pi (M_{\alpha n}^h - M_{\alpha n})/L$ for $\alpha = s,\eta$ and $n = 1,...,\bar{n}$.

The use of the distributions given in Eq. (5) of the Letter in the elementary $\alpha$-spin currents general expressions,
Eqs. (\ref{js}) and (\ref{jeta}), leads to the expression for the elementary $\alpha$-spin 
elementrary currents provided in Eq. (6) of the Letter. The parameter $j_{\alpha n}$ appearing
in that expression is for $\alpha = s$ and $\alpha = \eta$ given by,
\begin{equation}
j_{sn} = J_n \hspace{0.20cm}{\rm and}\hspace{0.20cm}j_{\eta n} = {4t\over \pi}\sqrt{1 + (nu)^2} - J_n \, ,
\label{Jalphan}
\end{equation}
respectively, where,
\begin{equation}
J_n = {t\over\pi}\int_{\pi y}^{\pi} dk\, 2\pi\rho (k) \left(1 - {2\over\pi}\arctan \left({\sin k\over n u}\right)\right)\sin k \, .
\label{Jn}
\end{equation}
Here $2\pi\rho (k)$ is defined by the integral equation, Eq. (A9) of Ref. \onlinecite{Carmelo_25},
with the distributions specific to the $\bar{n}$ states, Eq. (5) of the Letter.
\begin{table}
\begin{center}
\begin{tabular}{|c|c|c|c|c|c|c|c|} 
\hline
$c_0$ & $c_1$ & $c_2$ & $c_3$ & $c_4$ & $c_5$ \\
\hline
$1$ & ${3\over 2+\sqrt{3}}$ & ${11\over (2+\sqrt{3})^2}$ & ${41\over  (2+\sqrt{3})^3}$ & ${153\over  (2+\sqrt{3})^4}$ & ${571\over  (2+\sqrt{3})^5}$ \\
\hline
$1$ & $0.803848$ & $0.789764$ & $0.788753$ & $0.788681$ & $0.788676$ \\
\hline
\hline
$c_6$ & $c_7$ & $c_8$ & $c_9$ & $c_{10}$ & $c_{11}$ \\
\hline
${2131\over (2+\sqrt{3})^6}$ & ${7953\over  (2+\sqrt{3})^7}$ & ${29681\over  (2+\sqrt{3})^8}$ & ${110771\over  (2+\sqrt{3})^9}$ & ${413403\over  (2+\sqrt{3})^{10}}$ & ${1542841\over  (2+\sqrt{3})^{11}}$ \\
\hline
$0.788675$ & $0.788675$ & $0.788675$ & $0.788675$ & $0.788675$ & $0.788675$ \\
\hline
\end{tabular}
\caption{The constant $c_n$, Eq. (\ref{cn}), which obeys the exact sum rule, Eq. (\ref{csumrule}),
reads $1$ for $n=0$ and $c_n = \prod_{n'=1}^n \left({A_{n'}\over 2 + \sqrt{3}}\right)$ for $n=1,...,\infty$
where the coefficients $A_n$ are defined in Eqs. (\ref{Anxnnj}) and (\ref{xn}). In this
table values of $c_n$ are given for $n=0,1,...,11$. The table first and second lines give its
exact and approximate value, respectively. For $n=\infty$ it reads $c_{\infty} = 0.78867513459481...$.}
\label{table1}
\end{center}
\end{table}

At fixed $z = (L_{\eta} - L_s)/L$, the value of $y (z)$ depends on $u$.
Reciprocally, at fixed $\tau$-band momentum-rapidity $k = \pi y$
that separates the unoccupied interval $k\in [-\pi, \pi y]$ from the occupied 
interval $k\in [\pi y,\pi]$ of the $\tau$-band momentum-rapidity distribution
${\bar{N}}_{\tau} (k)$ in Eq. (5) of the Letter, $z (y) = (L_{\eta} - L_s)/L$ also
depends on $u$. Such a dependence on $u$ and that of the $\alpha n$-band numbers 
$M_{\alpha n}$ and $M_{\alpha n}^h$ for $\alpha =s,\eta$ and $n = 1,...,\bar{n}$ 
associated with $q_{\alpha n}^0 = {\pi \over L}(M_{\alpha n}^h - M_{\alpha n})$
is fully determined by the Bethe-ansatz equations, 
Eqs. (A5)-(A7) of Ref. \onlinecite{Carmelo_25}. This is achieved by using in these
equations the $\tau$-band momentum rapidity-variable distribution ${\bar{N}}_{\tau} (k)$ 
and $\alpha n$-band rapidity-variable distributions $\bar{M}_{n} (\Lambda)$ given in Eq. (5) of the Letter. Solution of the equations provides the 
corresponding needed dependence on $u=U/4t$ of the above quantities.

We then find that the ratios $2S_{\alpha}/L$, $M_{\alpha n}^h/L$, 
and $M_{\alpha n}/L$ where $2S_{\alpha}/L = M_{\alpha \bar{n}}^h/L$
are both for $\alpha =s$ and $\alpha = \eta$ given by,
\begin{eqnarray}
{2S_{\alpha}\over L} & = & {e^{-\bar{n}\ln (2+\sqrt{3})}\over c_{\bar{n}}}
\Bigl({1\over 2}(1 - \iota_{\alpha} z) + \sum_{n'=1}^{\bar{n}}2n' {Q_{\alpha n'}^0\over\pi}\Bigr)
\nonumber \\
{M_{\alpha n}^h\over L} & = & {c_{\bar{n}-n}\over c_{\bar{n}}}e^{-n\ln (2+\sqrt{3})} 
\Bigl({1\over 2}(1 - \iota_{\alpha} z) + \sum_{n'=1}^{\bar{n}}2n' {Q_{\alpha n'}^0\over\pi}\Bigr)
- \left({Q_{\alpha n}^0\over\pi}-{q_{\alpha n}^0\over\pi}\right)
\hspace{0.20cm}{\rm for}\hspace{0.20cm}n = 1,...,\bar{n}
\nonumber \\
{M_{\alpha n}\over L} & = & {c_{\bar{n}-n}\over c_{\bar{n}}}e^{-n\ln (2+\sqrt{3})} 
\Bigl({1\over 2}(1 - \iota_{\alpha} z) + \sum_{n'=1}^{\bar{n}}2n' {Q_{\alpha n'}^0\over\pi}\Bigr)
- {Q_{\alpha n}^0\over\pi}\hspace{0.20cm}{\rm for}\hspace{0.20cm}n = 1,...,\bar{n} \, ,
\label{NNSaG}
\end{eqnarray}
where $\iota_{\alpha} = \pm 1$ is defined in Eq. (\ref{PeffUalpha}).
\begin{figure*}
\includegraphics[width=0.495\textwidth]{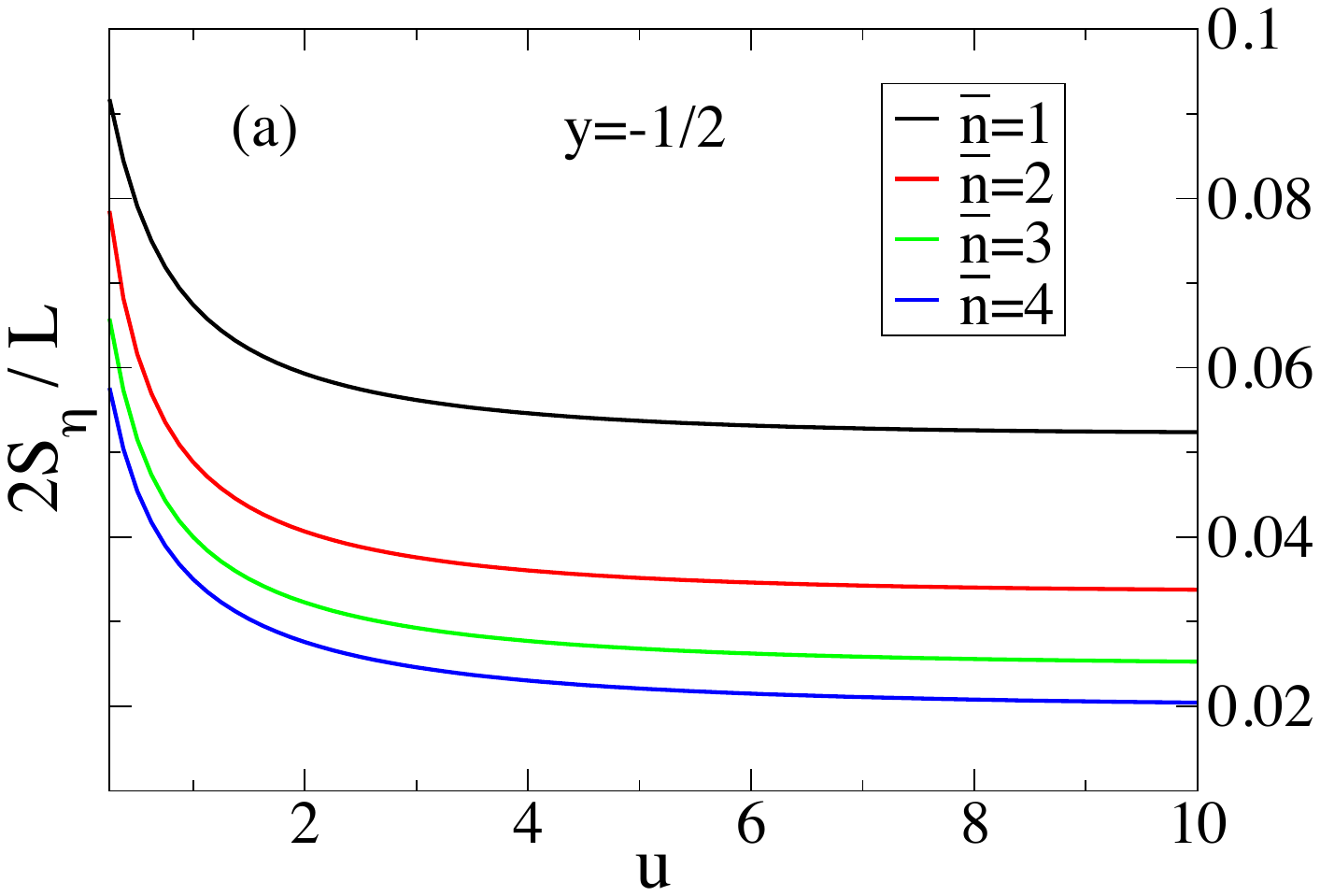}
\includegraphics[width=0.495\textwidth]{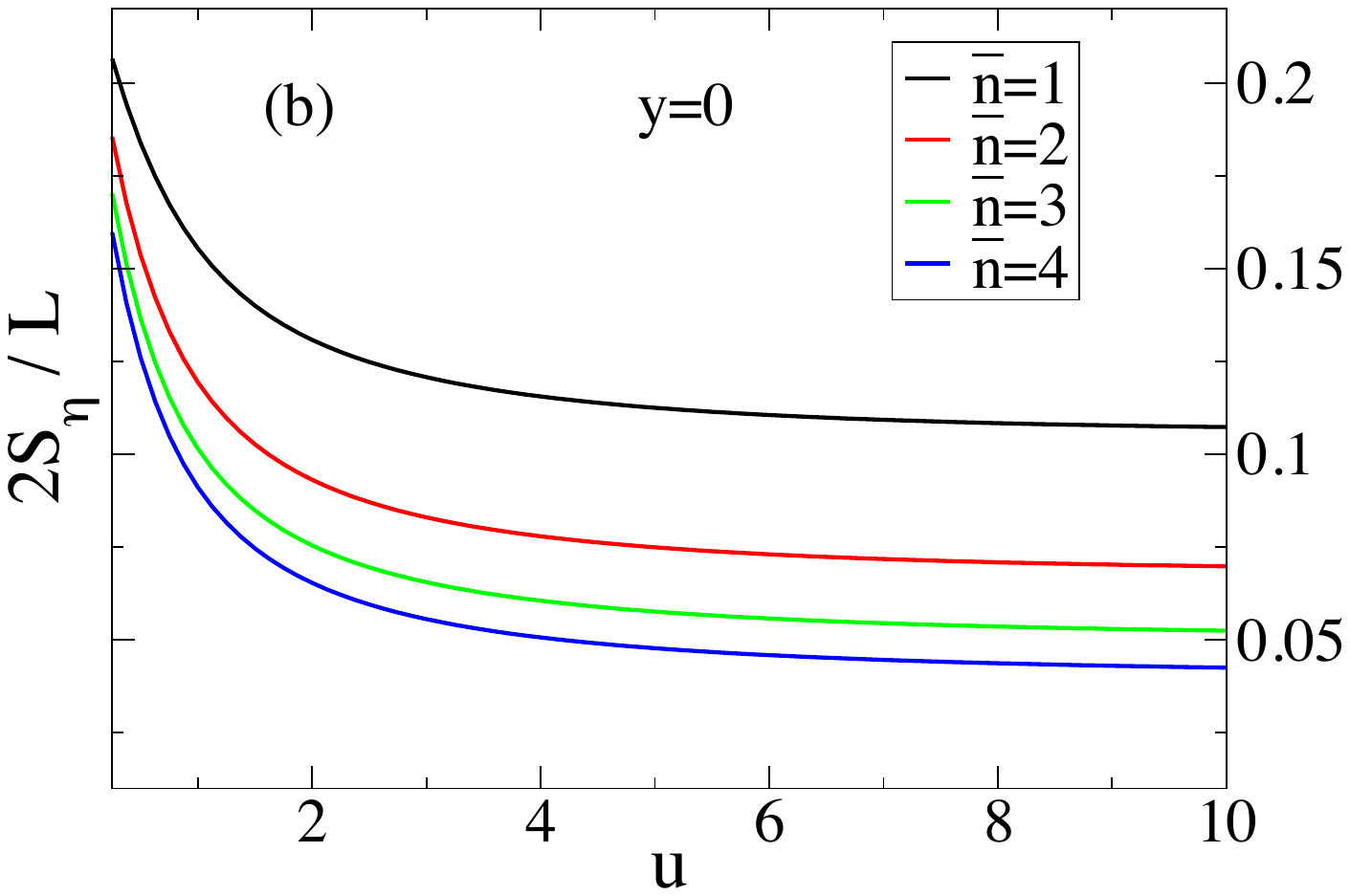}
\includegraphics[width=0.495\textwidth]{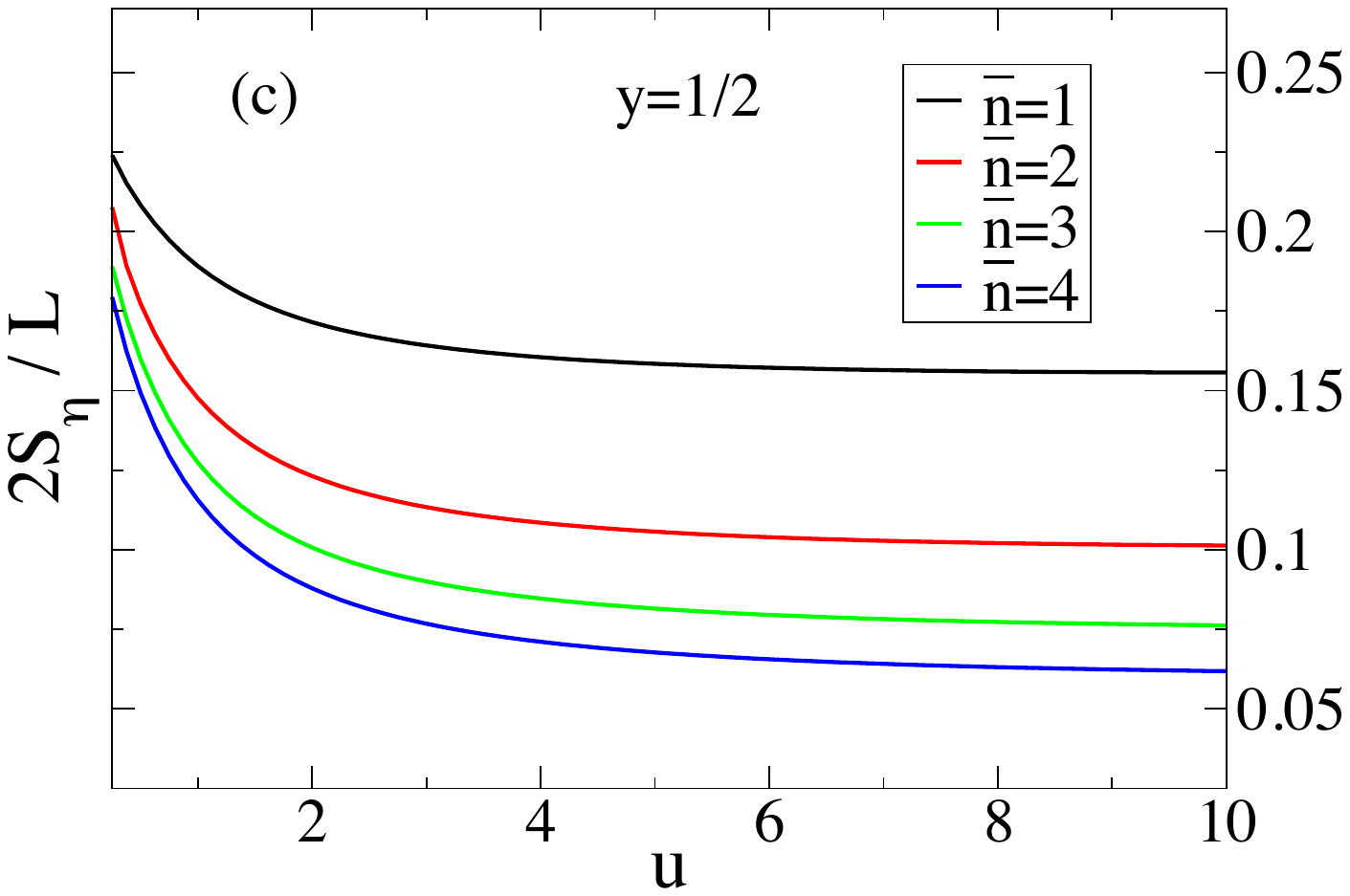}
\includegraphics[width=0.495\textwidth]{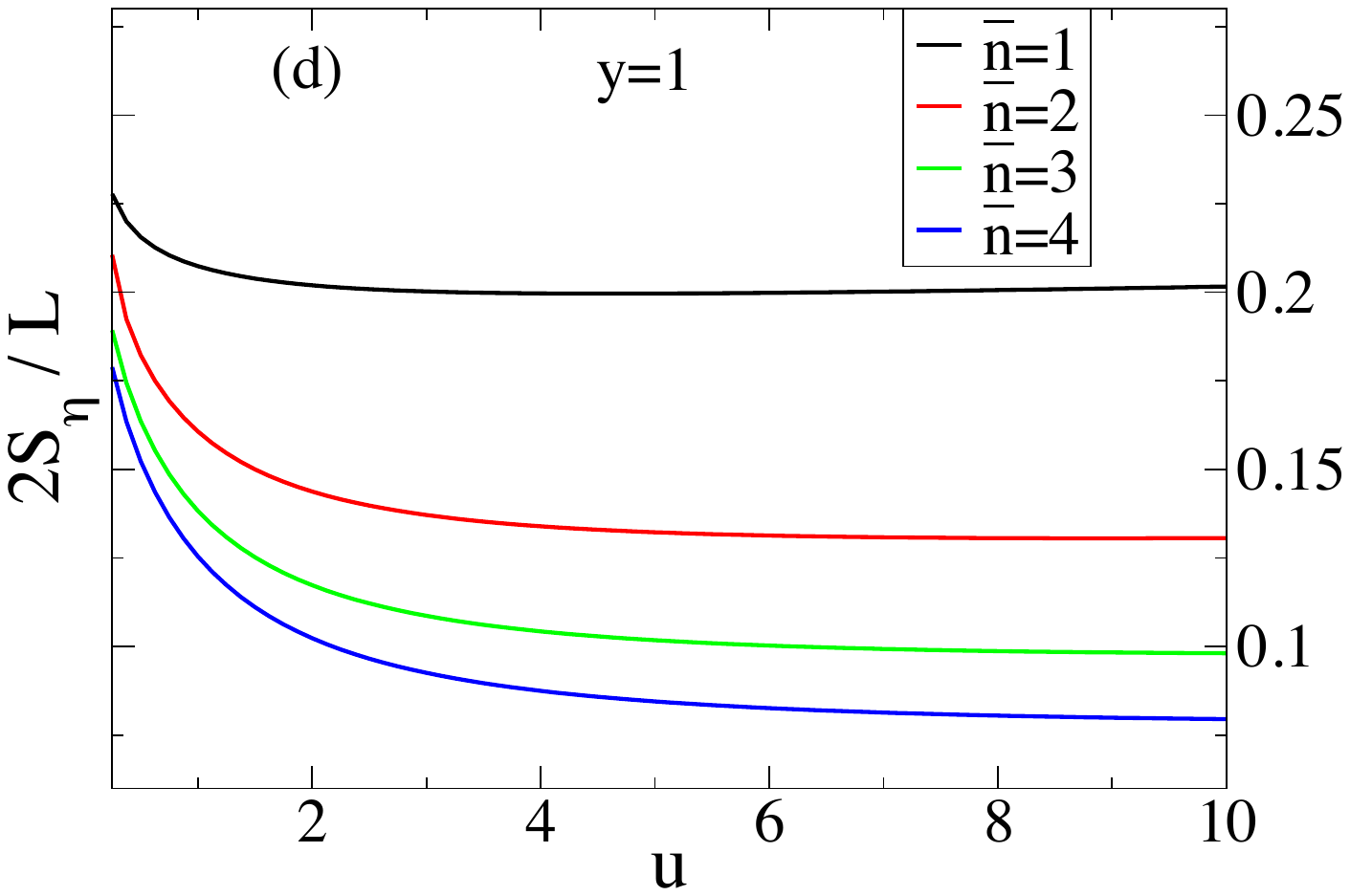}
\caption{The concentration of charge carriers of $\bar{n}$-states given in Eq. (\ref{NNSaG}) for $\alpha = \eta$
plotted as a function of $u$ for $\bar{n}=1,2,3,4$ and (a) $y = -1/2$,
(b) $y = 0$, (c) $y = 1/2$, and (d) $y = 1$.}
\label{figure1_CM_PRL}
\end{figure*}

The constant $c_n$ that in Eq. (\ref{NNSaG}) appears for $n$ given by $\bar{n}$ and $\bar{n}-n$, reads,
\begin{equation}
c_0 = 1 \hspace{0.20cm}{\rm and}\hspace{0.20cm}
c_n = \prod_{n'=1}^n\left({A_{n'}\over 2+\sqrt{3}}\right)
\hspace{0.20cm}{\rm for}\hspace{0.20cm}n=1,...,\infty \, ,
\label{cn}
\end{equation}
where,
\begin{equation}
A_{n} = {x_{n+1}\over x_n} 
\hspace{0.20cm}{\rm such}\hspace{0.20cm}{\rm that}\hspace{0.20cm}A_{n+1} - A_n = {2\over x_{n+1}\,x_n} \, ,
\label{Anxnnj}
\end{equation}
and the coefficients $x_n$ obey the following recursive relation,
\begin{equation}
x_n = 1 + \sum_{n'=1}^{n-1}2 (n-n') x_{n'} = \prod_{n'=1}^{n-1} A_{n-n'} 
= \prod_{n'=1}^{n-1} A_{n'} = c_{n-1}\,e^{(n-1)\ln (2+\sqrt{3})} \, ,
\label{xn}
\end{equation}
with boundary condition $x_1 = 1$, the use of which provides their general expression for $n = 1,...,\infty$.

The expression of the $n=1,...,\bar{n}$ momenta $Q_{\alpha n}^0$ also appearing in Eq. (\ref{NNSaG})
involves sums of the $n=1,...,\bar{n}$ $\alpha n$-band separation momentum values 
$q_{\alpha n}^0$. It also obeys a recursive relation,
\begin{equation}
Q_{\alpha n}^0 = q_{\alpha n}^0+ \sum_{n'=n+1}^{\bar{n}}2 (n'-n) Q_{\alpha n'}^0\hspace{0.20cm}{\rm for}\hspace{0.20cm}
n = 1,...,\bar{n} \, ,
\label{f0n}
\end{equation}
with the boundary condition $Q_{\alpha \bar{n}}^0 = q_{\alpha \bar{n}}^0$.
The separate momentum values $q_{\alpha n}^0$ such that $\Lambda_{\alpha n} (q_{\alpha n}^0 + q^{\Delta}) = 0$
only vanish for $u\rightarrow\infty$ and increase on decreasing $u$. 
This is consistent with the values of the numbers $M_{\alpha n}$ and $M_{\alpha n}^h$ 
in $q_{\alpha n}^0 = {\pi \over L}(M_{\alpha n}^h - M_{\alpha n})$ also changing
on changing $u$. Since different values for the numbers $M_{\alpha n}$ and $M_{\alpha n}^h$
refer to different energy eigenstates, the $\bar{n}$ states are a different energy eigenstate
for each value of $u$.

The constant $c_n$, Eq. (\ref{cn}), obeys at $n = \bar{n}$ the exact sum rule,
\begin{equation}
c_{\bar{n}} = e^{-\bar{n}\ln (2+\sqrt{3})} +
\sum_{n=1}^{\bar{n}}2n\,c_{\bar{n}-n}\,e^{-n\ln (2+\sqrt{3})} \, .
\label{csumrule}
\end{equation}

For $\bar{n}$-states, the concentration of $\alpha$-spin carriers $N_{\alpha}/L = 2S_{\alpha}/L$ given in Eq. (\ref{NNSaG})
decreases on increasing $u$, having the following limiting values,
\begin{eqnarray}
\lim_{u\rightarrow 0}{2S_{\alpha}\over L} & = & \lim_{u\rightarrow 0}{N_{\alpha}\over L} = {1\over 4}(1 - \iota_{\alpha} z)
\nonumber \\
\lim_{u\rightarrow\infty}{2S_{\alpha}\over L} & = & \lim_{u\rightarrow\infty}{N_{\alpha}\over L} = {(1 - \iota_{\alpha} z)\over 2c_{\bar{n}}(2+\sqrt{3})^{\bar{n}}} 
= {(1 - \iota_{\alpha} z)\,e^{-\bar{n}\ln (2+\sqrt{3})}\over 2c_{\bar{n}}}
\hspace{0.20cm}{\rm for}\hspace{0.20cm}\alpha = s,\eta  \, .
\label{2Slimits}
\end{eqnarray}
It is plotted in the case of charge carriers in Fig. \ref{figure1_CM_PRL} as a function of $u$ for 
$\bar{n}=1,2,3,4$ and (a) $y = -1/2$, (b) $y = 0$, (c) $y = 1/2$, and (d) $y = 1$. While $z (0) = 0$ and $z (1) = 1$,
$z = (L_{\eta} - L_s)/L$ is for $y=\pm 1/2$ closer to $\pm 1/2$ as $u$ increases. 
For instance, at $u=U/4t=1$ we have that $z (-1/2) \approx - 0.57$ for $\bar{n} = 1$, $z (-1/2) \approx - 0.58$ 
for $\bar{n} = 2,3,4$, and $z (1/2) \approx 0.63$ for $\bar{n} = 1,2,3,4$. The concentration of spin carriers
$2S_{s}/L$ has the same type of $u$ dependence.

At fixed $u$, the concentration of $\alpha$-spin carriers $N_{\alpha}/L=2S_{\alpha}/L$ decreases on increasing 
$\bar{n}$. There are two cases:\\

(1) When the concentration of $u\rightarrow\infty$ $\alpha$-spin carriers, $\lim_{u\rightarrow\infty}2S_{\alpha}/L = 
[(1 - \iota_{\alpha} z)\,e^{-\bar{n}\ln (2+\sqrt{3})}]/2c_{\bar{n}}$, is finite, the number $\bar{n}$ is finite for $u>0$
and the absolute value of the corresponding $\alpha$-spin elementary currents in Eq. (6) of the Letter
of such $\bar{n}$ states is finite for $u>0$.\\

(2) When the number of $u\rightarrow\infty$ $\alpha$-spin carriers, $\lim_{u\rightarrow\infty}2S_{\alpha}$, is finite, 
$2S_{\alpha} = 1,2,3,...$, and thus their concentration, $\lim_{u\rightarrow\infty}2S_{\alpha}/L = 
[(1 - \iota_{\alpha} z)\,e^{-\bar{n}\ln (2+\sqrt{3})}]/2c_{\bar{n}}$, vanishes
in the thermodynamic limit, $L\rightarrow\infty$, the exponential factor in it implies that
$\bar{n}\rightarrow\infty$. This $u\rightarrow\infty$ boundary condition is associated with $\bar{n}\rightarrow\infty$ for 
the whole $u>0$ range. That $\bar{n}\rightarrow\infty$ implies that the absolute value of the corresponding 
$\alpha$-spin elementary currents in Eq. (6) of the Letter of such $\bar{n}\rightarrow\infty$ states
diverges.\\

{\bf V - The $\alpha$-spin elementary currents of other states of classes (B)}\\

Importantly, there is in the thermodynamic limit a finite density of states of class (B).
We considered all different types of states of class (B) with finite occupancies
in $\alpha n$ bands for $n=1,...,\bar{n}$ where $\bar{n}=1,...,\infty$. For instance, instead of 
$\alpha n$-band rapidity distributions for $n=1,...,\bar{n}$ with rapidity occupancies 
of $\alpha n$-holes for $\Lambda\in [-\infty,0]$ and $\alpha n$-pairs for $\Lambda\in [0,\infty]$, Eq. (5) of the Letter, 
we considered states the $\alpha n$-band momentum distributions of which have for $n=1,...,\bar{n}$ occupancies of 
$\alpha n$-holes for $q \in [-\pi L_{s n}/L,0]$ and of $\alpha n$-pairs for $q \in [0,\pi L_{s n}/L]$.
The use of the corresponding distributions in the general expressions, Eqs. (\ref{jsqj}) and (\ref{jetaqj}), 
of the $\alpha$-spin elementary currents gives,
\begin{equation}
 j_{s,\pm 1/2} = \pm t{L\over N_{\tau}}{1\over\pi}\int_{\pi z}^{\pi} dq {\sin k (q)\over 2\pi\rho (k (q))}
\pm {8\over U}\sum_{n=1}^{\bar{n}} {L\over M_{sn}^h}
 {1\over\pi^2}\int_{-{\pi L_{s n}\over L}}^0 
 {dq\over 2\pi\sigma_{sn} (\Lambda_{sn} (q))}\int_{\pi z}^{\pi} dq'
 {\sin k (q')\over 1 + \left({\Lambda_{sn} (q) - \sin k (q')\over nu}\right)^2} \, ,
 \label{jsqjn}
 \end{equation}
 and
 \begin{eqnarray}
 j_{\eta,\pm 1/2} & = & \mp t{L\over N_{\tau}}{1\over\pi}\int_{-\pi}^{\pi z}dq{\sin k (q)\over 2\pi\rho (k (q))}
 \pm 2t\sum_{n=1}^{\bar{n}} {L\over M_{\eta n}^h}
 {1\over\pi^2}\int_{-{\pi L_{\eta n}\over L}}^0 {dq\over 2\pi\sigma_{\eta n} (\Lambda_{\eta n}(q))}
 \nonumber \\
 & \times & \left(n\,{\rm Re}\Bigl\{{(\Lambda_{\eta n} (q) - i n u)\over\sqrt{1 - (\Lambda_{\eta n} (q) - i n u)^2}}\Bigr\}
 + {4t\over U} \int_{\pi z}^{\pi} dq'{\sin k (q')\over 1 + \left({\Lambda_{\eta n} (q) - \sin k (q')\over nu}\right)^2}\right) \, ,
 \label{jetaqjn}
 \end{eqnarray}
 respectively. Such $q\bar{n}$ states become $\bar{n}$ states in the limit of $u\rightarrow\infty$.
 We checked that for finite $u$ values the absolute values of their $\alpha$-spin elementary currents,
 Eqs. (\ref{jsqjn}) and (\ref{jetaqjn}), are smaller than those of $\bar{n}$ states, Eq. (6) of the Letter.

We also considered states of class (B) with rapidity occupancies similar to those given in Eq. (5) of the Letter
but with the $\Lambda$ values separating $\alpha n$-holes from $\alpha n$-pairs deviating a bit from
$\Lambda = 0$. Similarly, we considered states of class (B) with $\alpha n$-bands momentum occupancies similar 
to those of the $q\bar{n}$ states associated with the $\alpha$-spin elementary currents provided in 
Eqs. (\ref{jsqjn}) and (\ref{jetaqjn}) but with the $\alpha n$ momentum values $q$ separating $\alpha n$-holes from 
$\alpha n$-pairs deviating a bit from $q = 0$. Again, all such states of class (B) have $\alpha$-spin elementary currents smaller than those of the 
$\bar{n}$ states, Eq. (6) of the Letter.

Importantly, all such states of class (B) have excitation energy density spectra in function of $z\in [-1,1]$ similar to those 
plotted in Fig. 2 of the Letter, in that such a density never vanishes for $z>-1$.\\

{\bf VI - Einstein relation and the charge dc conductivity}\\

According to the Einstein relation, the charge dc conductivity $\sigma_{\eta} (T)$ 
is related to the charge diffusion constant $D_{\eta} (T)$ as 
$\sigma_{\eta} (T) = \chi_{\eta} (T) D_{\eta} (T)$. Here $\chi_{\eta} (T)$ is the uniform charge susceptibility.

By combining the expressions given in Eq. (3) of the Letter with Eq. (54) of Ref. \onlinecite{Carmelo_25}, we find that 
at the $h=\mu = 0$ point the charge dc conductivity $\sigma_{\eta} (T)$ can be written for all finite temperatures $T>0$ as,
\begin{equation}
\sigma_{\eta} (T) = {L\,\langle\vert j_{\eta,\pm 1/2}\vert^2\rangle_{m_{s z},T}
\over 8 v_{\eta {\rm LR}}\,f_{\eta} (T)} = {L\,\langle\vert j_{\eta,\pm 1/2}\vert^2\rangle_{0,T}
\over 8 v_{\eta {\rm LR}}\,f_{\eta} (T)} \, .
\label{Generalsigma}
\end{equation}
Here $\langle\vert j_{\eta,\pm 1/2}\vert^2\rangle_{m_{s z},T}$ is defined in Eq. (4) of the Letter,
$m_{s z} = 0$ at the $h=\mu = 0$ point, $v_{\eta {\rm LR}}$ is the charge Lieb-Robinson velocity \cite{Carmelo_25}, 
and $f_{\eta} (T)$ is the second derivative of the 
free energy density with respect to $m_{\eta z}$ at $m_{\eta z} = 0$.

In the $k_B T/\Delta_{\eta} \ll 1$ regime where $\Delta_{\eta}$ is the Mott-Hubbard gap, the
charge carriers are $\eta$-spin triplet pairs \cite{Carmelo_24}. They correspond to two unpaired physical
$\eta$-spins with the same projection $+1/2$ or $-1/2$ and the two $\tau$-holes of which that describe their 
translational degrees of freedom have $\tau$-band momentum values that differ by $2\pi/L$.
The static mass $M_{\eta}$ and transport mass $M_t$ of the $\eta$-spin triplet pairs, the expressions 
of which are known \cite{Carmelo_25,Carmelo_24}, are related as $M_t = M_{\eta}/2$.

The expression of the dc charge conductivity, Eq. (\ref{Generalsigma}), 
simplifies in the $k_B T/\Delta_{\eta} \ll 1$ regime to \cite{Carmelo_25,Carmelo_24},
\begin{equation}
\sigma_{\eta} (T) = \left({1\over 4\pi M_t k_B T}\right)^{1/2} \, ,
\label{chisigma}
\end{equation}
so that $\sigma_{\eta}\rightarrow \infty$ in the $T\rightarrow 0$ limit. 

This though does not apply {\it at} zero temperature, at which the dc charge conductivity is up to finite frequencies $\omega$
just above $\Delta_{\eta}$ given by,
\begin{eqnarray} 
\sigma_{\eta}  (\omega) & = & 0 \hspace{0.20cm}{\rm for}\hspace{0.20cm} \omega < \Delta_{\eta}
\hspace{0.20cm}{\rm and}\hspace{0.20cm}T=0
\nonumber \\
& \sim & \left({\omega - \Delta_{\eta}\over 4\pi M_t\Delta_{\eta}^2}\right)^{1/2} \propto (\omega - \Delta_{\eta})^{1/2} 
 \hspace{0.20cm}{\rm for}\hspace{0.20cm}(\omega - \Delta_{\eta})/\Delta_{\eta} \ll 1
\hspace{0.20cm}{\rm and}\hspace{0.20cm}T=0 \, ,
\label{sigmaT0}
\end{eqnarray}
as provided in Eq. (29) of Ref. \onlinecite{Carmelo_04}. Hence $k_B T$ is 
replaced by $\Delta^2/(\omega - \Delta_{\eta})$ {\it at} $T=0$, where both $k_B T>0$
and $(\omega - \Delta_{\eta})\vert_{T=0}>0$ are very small but finite.\\ \\
{\bf References}
%%%%%%%%%%%%%%%%%%%%%%%%%%%%%%%%%%%%%%%%%%%%%%%%%%%%%%%%%%%%%%%%%%%%

\end{widetext}

\end{document}